\documentclass[a4paper,11pt]{article}
\pdfoutput=1

\usepackage{jheppub}
\usepackage{epsfig}

\usepackage[utf8]{inputenc}
\usepackage{color}
\usepackage{textcomp,color}
\usepackage{graphicx,subfigure}

\usepackage[normalem]{ulem}

\usepackage{changes}
\usepackage{lipsum}
\definechangesauthor[name={Per cusse}, color=orange]{per}

\title{Scattering of the asymmetric $\phi^6$ kinks from a $\mathcal{PT}$-symmetric perturbation: creation of multiple pairs of kink-antikink from phonons}

\author[a,1]{Danial Saadatmand,\note{Corresponding author.}}
\author[b]{Aliakbar Moradi Marjaneh}

\affiliation[a]{Department of Physics, University of Sistan and Baluchestan, Zahedan, Iran}
\affiliation[b]{Department of Physics, Quchan Branch, Islamic Azad University, Quchan, Iran}

\emailAdd{saadatmand.d@gmail.com}
\emailAdd{moradimarjaneh@gmail.com}

\abstract{Interaction of asymmetric $\phi^6$ kinks with a spatially localized $\mathcal{PT}$-symmetric perturbation is investigated numerically. It has been shown that when the kink (antikink) hits the defect from the gain side, a final velocity of the kink decreases (increases), while for the kink and antikink coming from the opposite direction their final velocities remain unchanged. It is also found that when the kink interacts with the defect from the gain side multiple pair of the kink-antikink are formed from small amplitude waves (phonons) in the final states depending on the initial velocity of the initial kink and parameter of the perturbation.}

\makeatletter
\def\@fpheader{\relax}
\makeatother

\begin{document}

\maketitle

\flushbottom

\section{Introduction}\label{sec:Introduction}
The standard concept of Hermiticity in quantum mechanics  was challenged by the great idea of  Bender and co-authors who studied a non-Hermitian Hamiltonians possess real spectra under the parity-time ($\mathcal{PT}$) symmetry condition, where parity-time means when one makes the spatial reflection $x\to -x$ and time reversal $t\to -t$, the  Hamiltonian is invariant\cite{Bender.PRL.1998,Bender.PRL.2002}. An enormous studies is devoted to develop physical systems with balanced gain and loss and such systems have been realized in optics \cite{Ruter,Guo,Regensburger,Regensburger1,Peng2014}, electronic circuits \cite{Schindler.PRA,Schindler.JPA,Bender.PRL.2013} and lasers \cite{Hodaei.science.2014,Longhi.PRA.2010}.

Many systems with $\mathcal{PT}$-symmetry have been attracting increased attention in different branches within physics because they can exhibit various interesting features such as spontaneous $\mathcal{PT}$-symmetry breaking \cite{Emery.JHEP,Ezawa.2021,Wen.2021}, associated ghost states \cite{Paul.2021,Bender.PRD.2005}, wave chaos \cite{West.PRL}, heat and mass transfer \cite{Ying Li et al}, supersymmetry \cite{Dorey.JPA}, microcavities \cite{Peng.Nat.Phys} and exciton-polariton condensates \cite{Lien.PRB,Gao.Nature,Chestnov.Sci.Rep}. 

The interaction of solitons and kinks with a localized defect has been extensively studied in the different contexts \cite{Khawaja.PRE.2021,Khawaja.EPL.2013,Javidan.PRE.2008,Danial.WRCM,Danial.Scripta,Kivshar.PLA.1988,Kivshar.Jpn.1987,Fei.SG.1992,Fei.phi4.1992}. The existence of the spectral walls and their properties are common phenomena in the dynamics of kinks in (1+1) dimensions with Bogomolnyi-Prasad-Sommerfield (BPS) preserving impurity \cite{Adam.PRL.2019,Adam.JHEP.2021}. In \cite{Javidan.PRE.2008,Danial.Scripta} different methods have been introduced to study the interaction of nonlinear Klein-Gordon solitons with defects. The collision of a fluxon with a localized defect which is a combination of a micro-short or micro-resistor with a dissipative inhomogeneity has been studied in \cite{Kivshar.PLA.1988,Kivshar.Jpn.1987}. One of the most interesting behaviour of the solitons during the interaction with defect is the quantum-behaviour of soliton which was first observed by Fei et al in \cite{Fei.SG.1992,Fei.phi4.1992}. The authors have shown that when a kink interacts with a potential well it may reflect back from a potential well if its initial velocity lies in some windows. This effect is due to the resonant energy exchange between the kink translational mode, its internal mode, and the impurity mode.

Two nonconservative perturbation examples of field
theories with a $\mathcal{PT}$-symmetric localized perturbation describing a defect with balanced gain and loss, namely, the
sine-Gordon (sG) and the $\phi^4$ models have been introduced by Kevrekidis \cite{kevrekidis2014variational}. Interaction of the $\phi^4$ kinks with and without internal mode excitation with a spatially localized $\mathcal{PT}$-symmetric defect was investigated numerically and analytically \cite{Danial.CNSNS.2015,Danial.JETP.2015}. It was demonstrated that the a noticeable internal mode is excited when kink passing through the defect from the gain side, while for the kink coming from the opposite direction the mode excitation is much weaker. The resonant interaction of the $\phi^4$ kink with a periodic $\mathcal{PT}$-symmetric perturbation was also observed in the frame of the continuum model and collective variable method \cite{Danial.CNSNS.2018}. The results found in the last paper indicate that when the kink interacts with the perturbation, the kink’s internal mode is excited with the amplitude varying in time quasiperiodically. The interaction of the moving kinks and breathers with the spatially localized $\mathcal{PT}$-symmetric perturbation was also investigated in the integrable sine-Gordon model \cite{Danial.PRE.2014}. It was shown that the only effect of the interaction of the sG kink with the defect is a kink phase shift. For the breather, the interaction was more interesting since the breather can split into subkinks depending on its parameters and also on the amplitude of the defect.

In contrast to the symmetrical kinks in the integrable sine-Gordon model, the symmetrical kinks in non-integrable double sine-Gordon and $\phi^4$ models have an internal vibrational mode. Some new phenomenons like entanglement \cite{,Alonso.JPB.2015}, formation of a bound state of two oscillons and the emergence of new resonance windows are explained based on internal mode in the latter models \cite{Campbell.dsG.1986, Gani.EPJC.2018,Gani.EPJC.2019,Alonso.PRD.2021,Alonso.CNSNS.2021}. Moreover, kinks in the higher order models, e.g., $\phi^6$ and $\phi^8$ models are anti-symmetric and do not support internal mode. For that reason kinks interaction in higher order models is more interesting and some phenomenons like resonant scattering structure, escape windows \cite{Dorey.PRL.2011,Gani.PRD.2020, Gani.EPJC.2021} or even the extreme values of the energy densities \cite{Moradi.JHEP.2017} depend on the order in which kinks collide. Many important results have also been obtained for topological field configurations in \cite{Bazeia.PLB.2019,Bazeia.EPJC.2021,Campos.JHEP.2021,Mohammadi.CNSNS.2019,Mohammadi.CNSNS.2020,weigel.PRD,weigel.Chapter,Moradi.CNSNS.2017,Malomed.CNSNS.2017,Peyravi.EPJB.2009,Peyravi.EPJB.2010,Askari.CSF.2020,Christov.PRL.2019,Zhong.JHEP.2020,Yan.PLB.2020}. Consequently, it is important to know how an asymmetric $\phi^6$ kink treats when it interacts with a $\mathcal{PT}$-symmetric perturbation. In the following we will show that the asymmetric nature of the kink in the $\phi^6$ model plays important role during the scattering from the $\mathcal{PT}$-symmetric defect.

Our paper is organized as follows. In section \ref{sec:Model} we briefly describe the spatially localized $\mathcal{PT}$-symmetric inhomogeneity into the (1+1)-dimensional $\phi^6$ model and presents its topologically non-trivial solutions—kinks and antikinks. The numerical scheme of the equation of motion and methods are illustrated in section \ref{sec:Results}. In section \ref{Sec:CollectVar} a one-degree of freedom collective variable method is introduced to show some feathers of the kink dynamics in the considered system. We report the results of the kink-defect interaction in section \ref{Sec:Kink-defect} and  kink-antikink-defect in section \ref{Sec:Kink-anitkink}. In section \ref{Sec:Conclusions} we give the conclusion and some future directions.

\section{The model}\label{sec:Model}

Here we introduce the $\phi^6$ model in $(1+1)$-dimensional space-time with the following Lagrangian density
\begin{equation}\label{eq:Largangian}
\mathcal{L} = \frac{1}{2}\left(\frac{\partial\phi}{\partial t}\right)^2 - \frac{1}{2}\left(\frac{\partial\phi}{\partial x}\right)^2 - U(\phi),
\end{equation}
in which $\phi(x,t)$ is a real scalar field. The on-site potential $U(\phi)$ has the form
\begin{equation}\label{eq:potential}
U(\phi)=\frac{1}{2}\phi^2(1-\phi^2)^2.
\end{equation}

\begin{figure*}[h!]
\begin{center}
  \centering
  \subfigure[]{\includegraphics[width=0.45
 \textwidth]{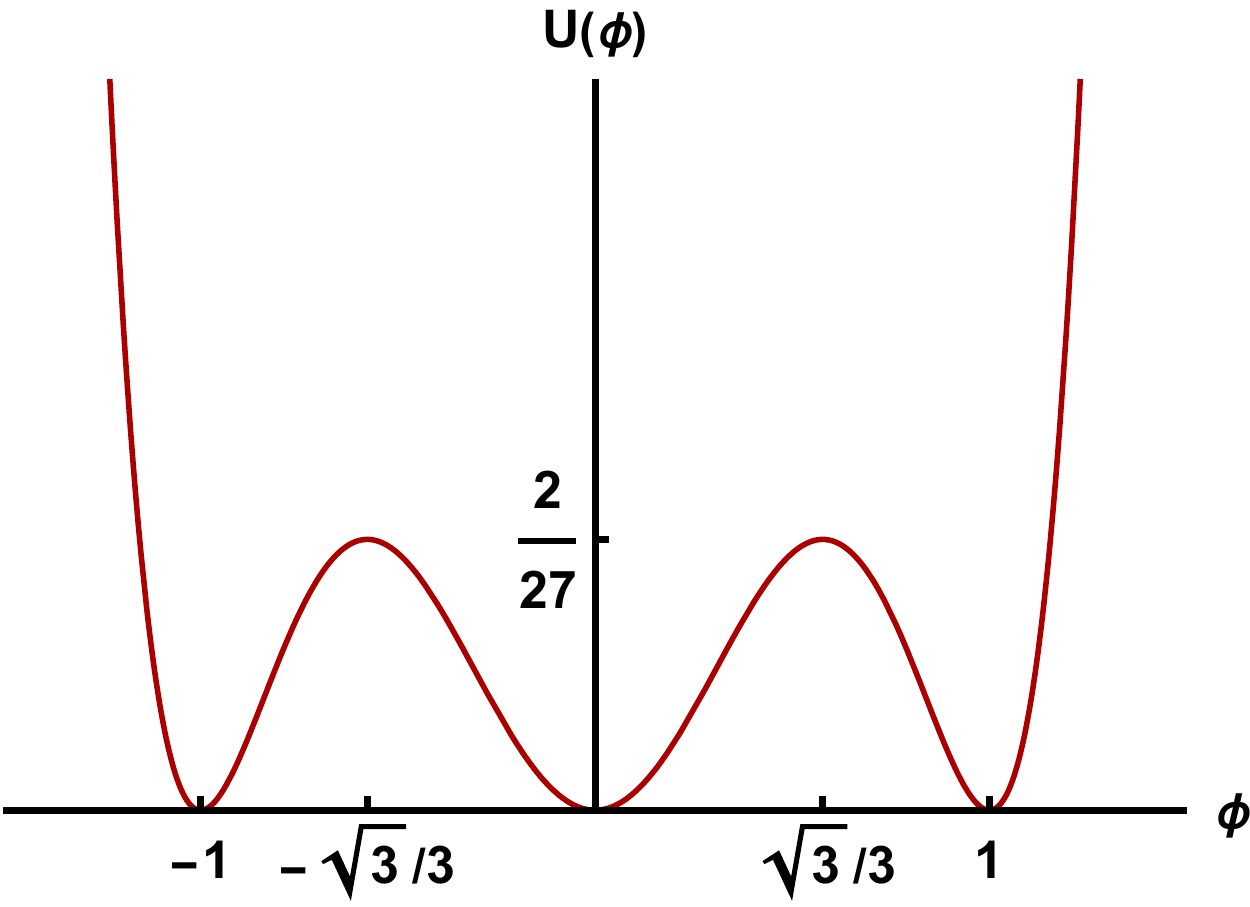}\label{fig:potential}}
  \subfigure[]{\includegraphics[width=0.45
 \textwidth]{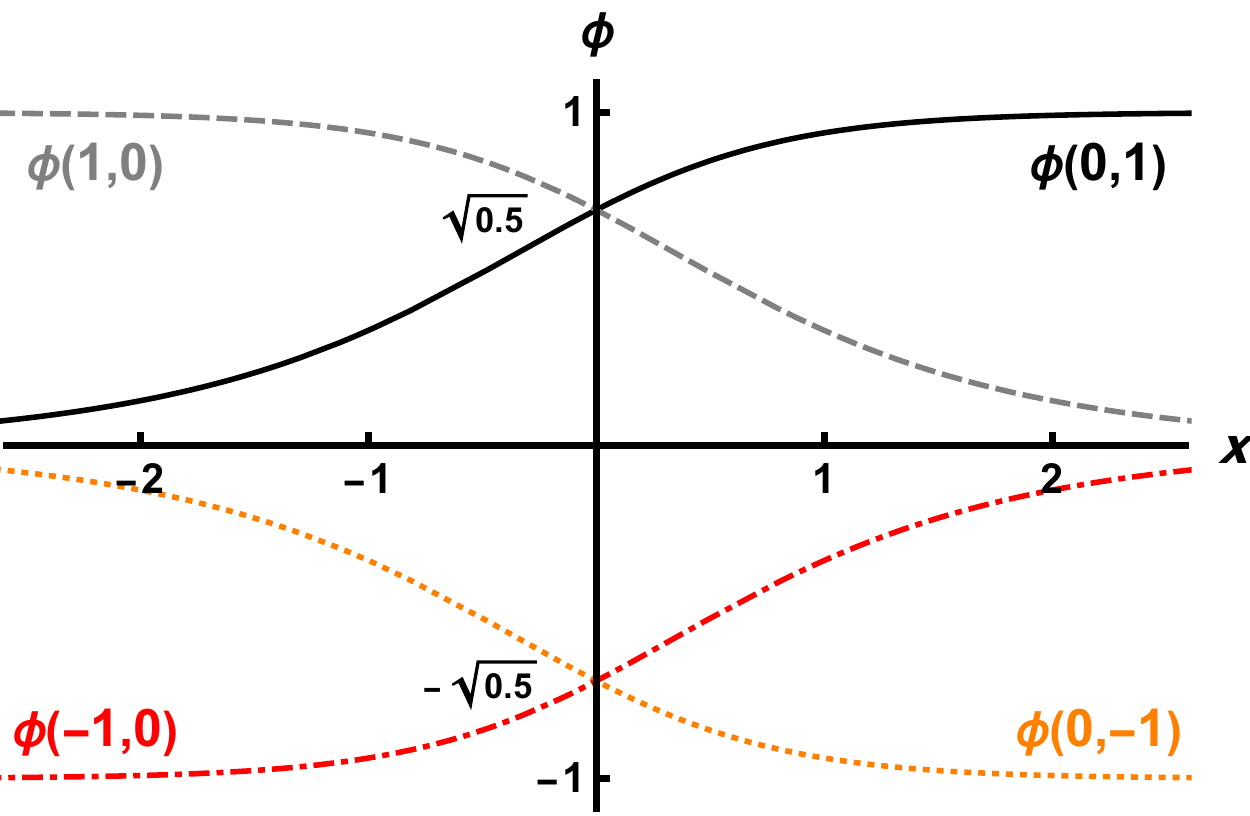}\label{fig:kinks}}
\\
  \caption{(a) The potential \eqref{eq:potential} of the $\phi^6$ model. (b) Kinks and antikinks solutions in the different cases, $a=\tanh^{-1}\left(\frac{1}{2}\right)\approx 0.55$ shows the center of kinks and antikinks.}
  \label{fig:PotentialKinks}
\end{center}
\end{figure*}
Total energy corresponding to the Lagrangian \eqref{eq:Largangian} is
\begin{equation}\label{eq:energy}
E[\phi]=\int_{-\infty}^{+\infty}\left[\frac{1}{2}\left(\frac{\partial\phi}{\partial t}\right)^2 + \frac{1}{2}\left(\frac{\partial\phi}{\partial x}\right)^2 + U(\phi)\right]dx.
\end{equation}
The Lagrangian \eqref{eq:Largangian} defines the following equation of motion
\begin{equation}\label{eq:eom}
\frac{\partial^2\phi}{\partial t^2} - \frac{\partial^2\phi}{\partial x^2} + \frac{dU}{d\phi}=0.
\end{equation}
The potential \eqref{eq:potential} has three minima of the same depth and thus, equation \eqref{eq:eom} possesses three static vacua solutions: $\bar{\phi}_1=-1$, $\bar{\phi}_2=0$, and $\bar{\phi}_3=1$, with equal energies $U(\bar{\phi}_1)=U(\bar{\phi}_2)=U(\bar{\phi}_3)=0$, see Fig.~\ref{fig:potential}. Considered model supports the kinks, in other words topological solitons, $\phi_\mathrm{K}(x)$ interpolating between neighboring vacua.

The notation $\phi_\mathrm{K}(x)$ is introduced for the kink belonging to the topological sector $(\bar{\phi}_i,\bar{\phi}_j)$ if $\lim\limits_{x\to-\infty} \phi_\mathrm{K}(x)=\bar{\phi}_i$ and $\lim\limits_{x\to+\infty} \phi_\mathrm{K}(x)=\bar{\phi}_j$. This kink is also denoted as $\phi_{(\bar{\phi}_i,\bar{\phi}_j)}(x)$ instead of $\phi_\mathrm{K}(x)$. The static kink solution of the $\phi^6$ model can be found analytically by solving Eq.~\eqref{eq:eom}. The result reads
\begin{equation}
\frac{d^2\phi}{dx^2}=\frac{dU}{d\phi}.
\end{equation}
This equation can be reduced to the first order ordinary differential equation
\begin{equation}
\frac{d\phi}{dx}=\pm\sqrt{2U(\phi)}.
\end{equation}
The on-site potential \eqref{eq:potential} has three minima, that is why the model supports two kinks and two antikinks, as shown in Fig.~\ref{fig:kinks}. The kinks are located in the topological sectors $(-1,0)$ and $(0,1)$, while the antikinks belong to the the sectors $(0,-1)$ and $(1,0)$. Below, in some cases, for brevity we use the term ``kink'' for both kinks and antikinks.

All static kinks and antikinks supported by the $\phi^6$ model can be expressed as:
\begin{equation}\label{eq:kinks1}
\phi_{(0,1)}(x) =\sqrt{\frac{1+\tanh x}{2}},
\quad
\phi_{(1,0)}(x) = \sqrt{\frac{1-\tanh x}{2}},
\end{equation}
\begin{equation}\label{eq:kinks2}
\phi_{(-1,0)}(x) = -\sqrt{\frac{1-\tanh x}{2}},
\quad
\phi_{(0,-1)}(x) = -\sqrt{\frac{1+\tanh x}{2}}.
\end{equation}
It is easy to demonstrate that the $\phi^6$ kinks are  asymmetric because they have different asymptotics bahaviours for the left and right tails. For example, for the kink $\phi_{(0,1)}(x)$ at $|x|\gg 1$ we have
\begin{equation}\label{eq:asymptotics1}
\phi_{(0,1)}(x)\sim e^x, \quad x\to-\infty,
\end{equation}
\begin{equation}\label{eq:asymptotics2}
\phi_{(0,1)}(x)\sim 1-\frac{1}{2}e^{-2x}, \quad x\to+\infty,
\end{equation}
while for the kink $\phi_{(1,0)}(x)$ the results are
\begin{equation}\label{eq:asymptotics1a}
\phi_{(1,0)}(x)\sim 1-\frac{1}{2}e^{2x}, \quad x\to-\infty,
\end{equation}
\begin{equation}\label{eq:asymptotics2a}
\phi_{(1,0)}(x)\sim e^{-x}, \quad x\to+\infty.
\end{equation}
As it will be shown, the exponential behaviour of the kink at infinities noticeably affect the kink dynamics during the interaction with a $\mathcal{PT}$-symmetric defect and it is important from which tail the kink hits the defect.

The mass of kink (or antikink) is $M_\mathrm{K}=\displaystyle\frac{1}{4}$, which can be found from substitution of Eqs.~\eqref{eq:kinks1} or \eqref{eq:kinks2} into the energy functional \eqref{eq:energy}. A moving kink (or antikink) can be derived from Eqs.~\eqref{eq:kinks1} and \eqref{eq:kinks2} with the use of the Lorentz boost, e.g.,
\begin{equation}
\phi_{(0,1)}(x,t) = \phi_{(0,1)}(\alpha(x-Vt)),
\end{equation}
and similarly for other kinks. Here $V$ is the kink velocity and $\alpha = 1/\sqrt{1-V^2}$.

\begin{figure}[h]
\centering
\includegraphics[width=0.65\textwidth]{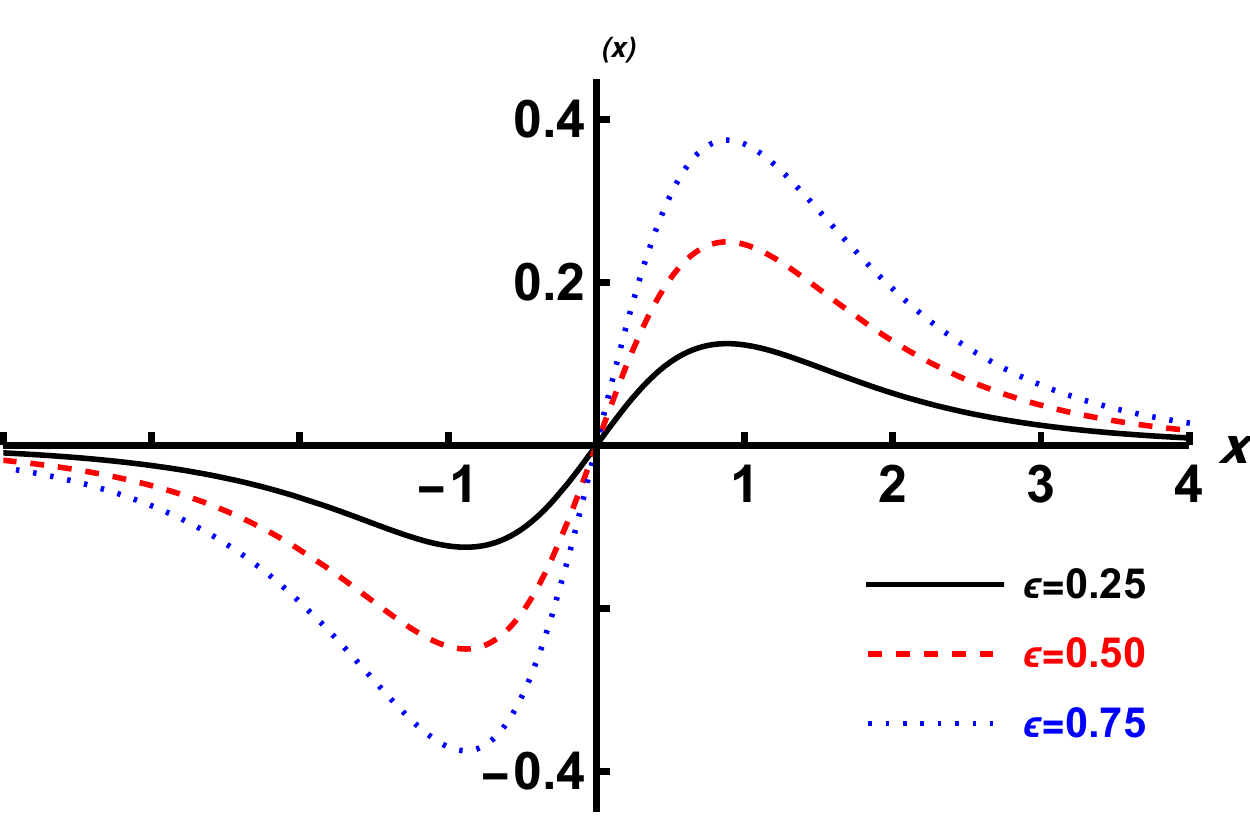} 
\caption{The profile of $\mathcal{PT}$-symmetric  defect \eqref{defect} for three different amplitudes, $\epsilon=0.25$ (thick-black), $\epsilon=0.5$ (red-dashed), and $\epsilon=0.75$ (blue-dotted) and fixed inverse width $\beta=1$.} 
\label{fig:perturbation}
\end{figure}

In order to study the scattering of moving $\phi^6$ kinks from the $\mathcal{PT}$-symmetric defect, the perturbation term is added to the right-hand side of equation (\ref{eq:eom}):
\begin{equation}\label{eq:eom1}
\frac{\partial^2\phi}{\partial t^2} - \frac{\partial^2\phi}{\partial x^2} + \frac{dU}{d\phi}=\gamma(x)\frac{\partial\phi}{\partial t},
\end{equation}
with
\begin{equation}\label{defect}
\gamma(x)=\epsilon\tanh(\beta x) {\rm sech}(\beta x),
 \end{equation}
where $\beta$ shows the defect inverse width and $\epsilon$ is the defect amplitude. From the physical point of view, Eq.~(\ref{eq:eom1}) describes an open system with balanced gain and loss. The profile of the defect has been plotted in Fig.~\ref{fig:perturbation} for the three different amplitudes, $\epsilon=0.25$, 0.5, and 0.75, and fixed inverse width, $\beta=1$.
\section{Numerical scheme}
\label{sec:Results}

Numerical integration of the equation of motion \eqref{eq:eom1} is based on its spatial discretization of the form
\begin{eqnarray}\label{Phi4discrete}
\frac{d^2\phi_n}{dt^2} - \frac{1}{h^2}(\phi_{n-1} -2\phi_{n}+\phi_{n+1}) 
&+\displaystyle\frac{1}{12h^2}(\phi_{n-2}-4\phi_{n-1} +6\phi_{n}-4\phi_{n+1}+\phi_{n+2})\nonumber\\
&+\phi_n(1-\phi_n^2)(1-3\phi_n^2)-\gamma_n \frac{d\phi_n}{dt}  = 0,
\end{eqnarray}
where $h$ is the lattice spacing, $n=0,\pm1,\pm2,...$, and $\phi_n(t)=\phi(nh,t)$. Note that the term $\phi_{xx}$ in Eq.~\eqref{Phi4discrete} is discretized with the accuracy $O(h^4)$ in order to reduce the effect of discreteness. For the integration of the equations of motion~\eqref{Phi4discrete}, an explicit St\"ormer method with the time step $\tau=0.005$ and the accuracy $O(\tau^4)$ is used.

In the numerical simulations presented in this section we use $5000$ points for the spatial grid, corresponding to the $x$ range from $-250$ to $250$ for $h=0.1$ and from $-125$ to $125$ for $h=0.05$. Hence the spatial boundaries are far away enough and cannot affect the numerical results. In addition, we also use the absorbing boundary conditions in order to prevent the small amplitude radiation reflected from the boundaries. 

\section {Collective variable method} \label{Sec:CollectVar}
In this section we use a one-degree of freedom collective variable model presented in \cite{kevrekidis2014variational} to describe the dynamics of the kink. The $\phi^6$ kink is effectively described by the one degree of freedom (kink's translational mode) particle of mass $M=\frac{1}{4}$, which is the mass of standing kink. The kink coordinate $X(t)$ (which in the unperturbed case is given
by $x_0+V_k t$ as a function of time $t$) can be found from the following equations
\begin{eqnarray}\label{Coll_var1}
    M\ddot{X}=\epsilon\dot{X}\int_{-\infty}^{\infty} [\phi^{\prime}
    (x-X)]^2\gamma(x)dx,
\end{eqnarray}
in which $\phi$ is the kink solution of the $\phi^6$ model which is given by Eqs.~(\ref{eq:kinks1}) and (\ref{eq:kinks2}).

A kink approaches the defect from the gain side always passes, while for the kink coming from the loss side it must have sufficient initial energy to pass through the defect. The kink critical velocity $V_{c}$ can be found with help of collective variable method. From Eq.~\eqref{Coll_var1} one can easily find
\begin{eqnarray}\label{Coll_CriticalV}
    M(\dot{X}-\dot{X_{0}})=\frac{\epsilon \delta_{k}^{2}}{8}\int_{-\infty}^{\infty}\int_{-X_0}^{X} \frac{\text{sech} ^{4}[\delta_{k}(x-X)]\gamma(x)dxdX}{(1\pm\tanh[\delta_{k}(x-X)])},
\end{eqnarray}
where the sign `$+$' is taken for the kinks $\phi_{(0,1)}$ and $\phi_{(0,-1)}$,  while the sign `$-$' corresponds to $\phi_{(1,0)}$ and $\phi_{(-1,0)}$. The value of integral in Eq.~\eqref{Coll_CriticalV} over the variable $x$ can be found numerically for the initial conditions $X_0=-20$, $X=0$ for the case when the parameters of the kink are $M=1/4$, $\beta=1.0$ and $\delta_{k}=1$ (which denotes kink moves with small velocity). After integrating over the collective variable $X$ in Eq.~\eqref{Coll_CriticalV} and setting $V_c=\dot{X_0}$ and $\dot{X}=0$, for kink critical velocity we have
\begin{equation}\label{CriticalV}
   V_{c}=0.78539\epsilon
\end{equation}
It should be mentioned that the critical velocity of the kink $\phi_{(0,1)}$ obtained with help of collective variable method is equal to the critical velocity of the antikink $\phi_{(1,0)}$ up to five decimal number. 

\section{Kink interaction with defect}\label{Sec:Kink-defect}
\begin{figure}[h]
\centering
\includegraphics[width=1.0\textwidth]{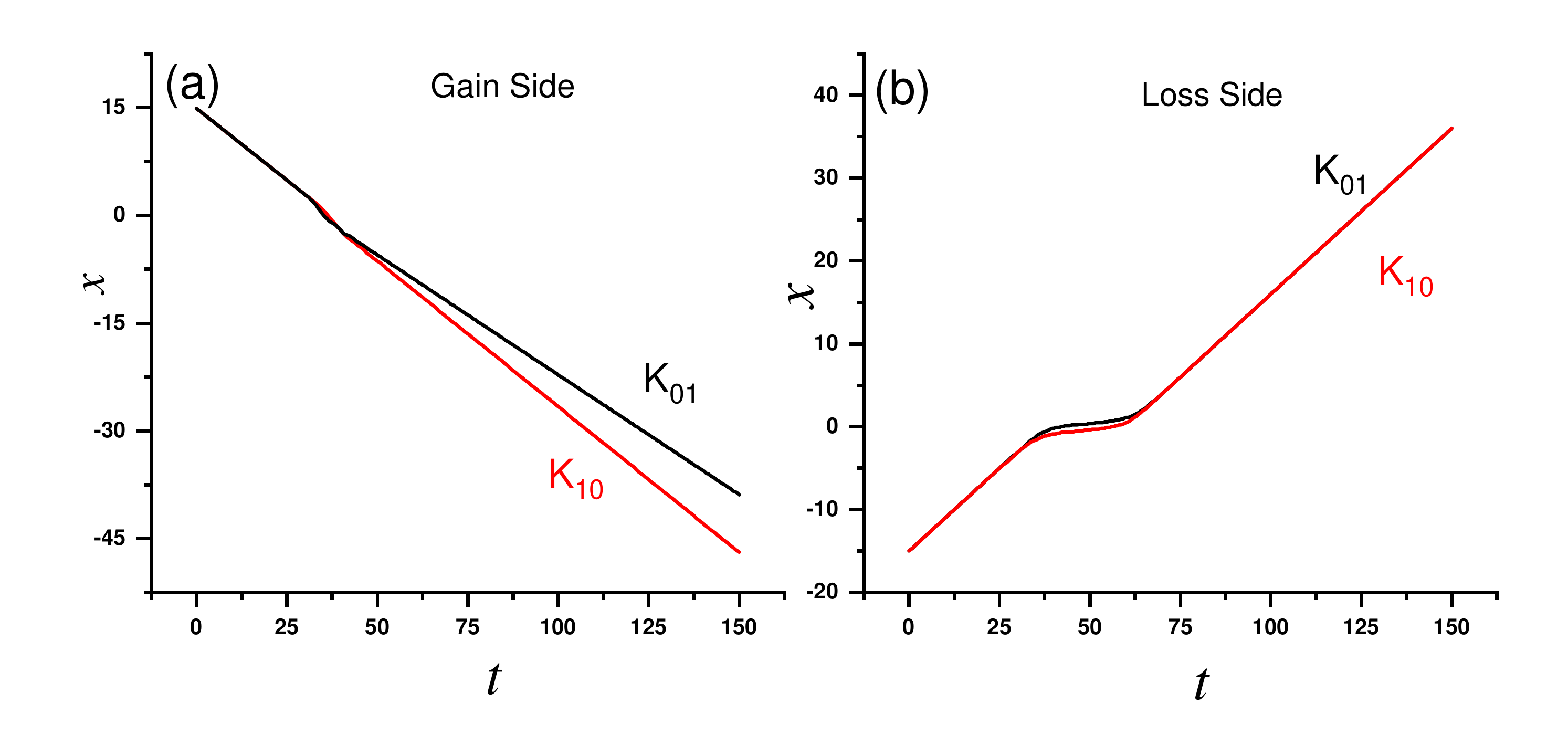} 
\caption{Kink position as the function of time with initial velocity $V=0.4$ for the case when the kink approaches the defect (a) from the gain side (b) from the loss side. The perturbation parameters are $\epsilon=1.2$ and $\beta=2.0$ in this case.} 
\label{fig:Trajectory2}
\end{figure}
Here we only investigate the interaction of the kink $\phi_{(0,1)}$ and antikink $\phi_{(1,0)}$ with a $\mathcal{PT}$-symmetric defect \eqref{defect}. One can easily found that the same results can be found for the kinks with the same initial velocities and initial positions belong to the topological sectors (0,-1) and (-1,0) (see ref.~\cite{Moradi.JHEP.2017}). 

The result of interaction of the kinks $\phi_{(0,1)}$ and $\phi_{(1,0)}$ with a $\mathcal{PT}$-symmetric defect are presented in Fig.~\eqref{fig:Trajectory2} for the case when kinks approach the defect with $\epsilon=1.2$ and $\beta=2.0$ (a) from the gain side (b) from the loss side. The figure shows kink position as the function of time for the kink initial velocity $V=0.4$. In (a) it is clearly seen that both kinks moving toward the defect from the gain side are first accelerated and then decelerated by the defect. It can be seen that  kink $\phi_{(1,0)}$ moves faster after the interaction with defect while kink $\phi_{(0,1)}$ becomes slower after passing through the defect. 
In Fig.~\eqref{fig:Trajectory2}~(b) kinks come from the loss side of the defect. In \cite{Danial.CNSNS.2015,Danial.PRE.2014} we have shown that two different scenarios are possible for the kink interaction with defect depending on initial velocity $V$ of the kinks. Here we only consider the kink with initial velocity $V=0.4$ which is above the kink critical velocity $V_c$. As a result, kink passes through the defect and it restores its initial velocity and then goes on to infinity. It can be seen when the kinks $\phi_{(0,1)}$ and $\phi_{(1,0)}$ come from the loss side the final velocity of the kinks are the same.
\begin{figure}[h]
\centering
\includegraphics[width=1.0\textwidth]{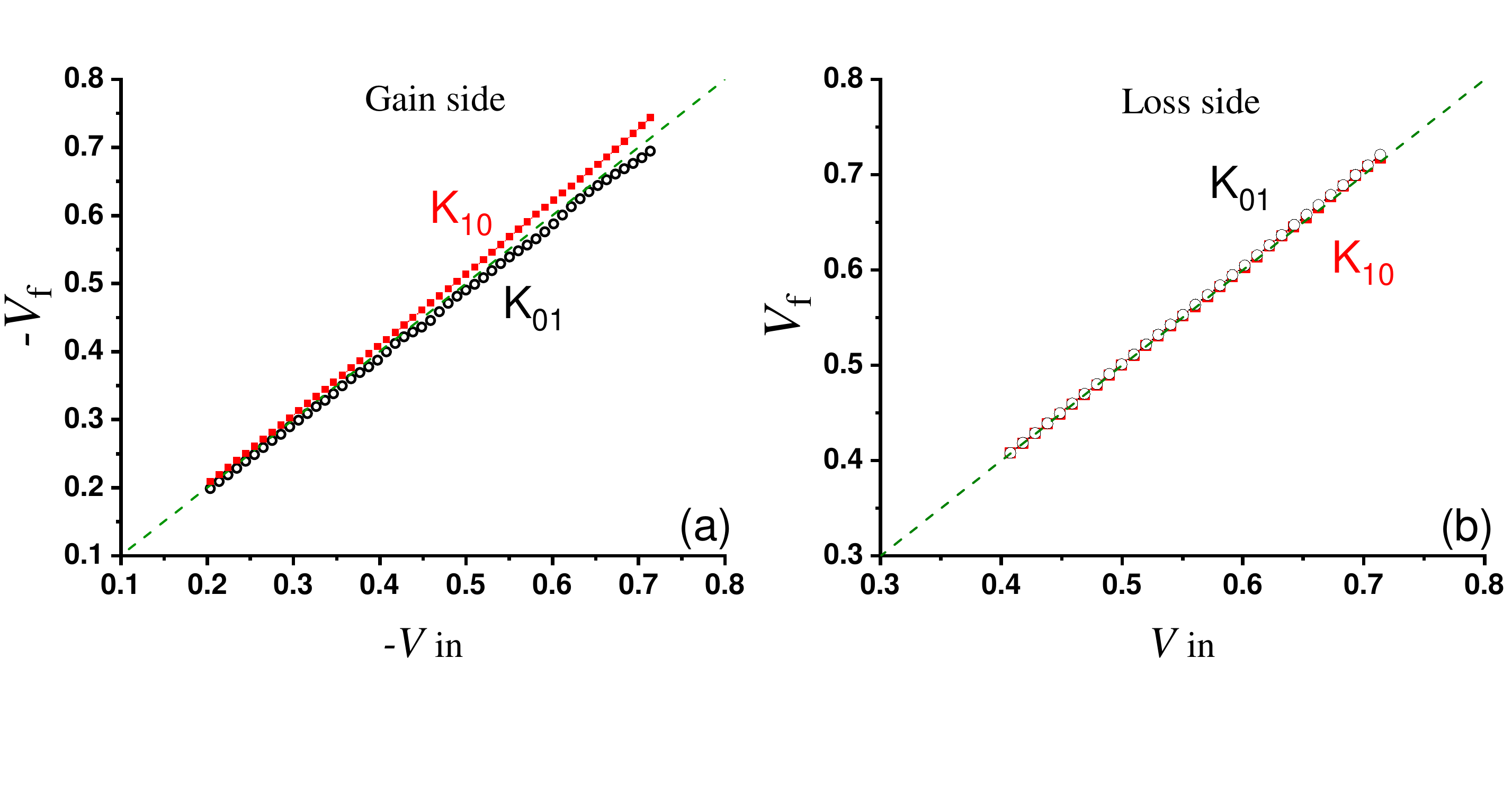} \caption{The final velocity of the kink $V_f$ after interaction with defect as a function of initial velocity $V_i$ when the kink approaches the defect (a) from the gain side (b) from the loss side. The perturbation parameters are $\epsilon=1.2$ and $\beta=2.0$ in this case.} 
\label{fig:FinalVelocity}
\end{figure}

In Fig.~\ref{fig:FinalVelocity} the final velocity of the kink $V_f$ after the interaction with a $\mathcal{PT}$-symmetric defect as the function of initial velocity $V_i$ is presented for the case when kink comes from (a) the gain side and (b) the loss side. The black-circle (red-square) line indicates the numerical results extracted for the kink $\phi_{(0,1)}$ ($\phi_{(1,0)}$). From Fig.~\eqref{fig:FinalVelocity}~(a) one can see when the kink $\phi_{(0,1)}$ ($\phi_{(1,0)}$) hits the defect from the gain side and passes through it, the final velocity of the kink decreases (increases) after the interaction with defect. The plot demonstrates that this reduction (increase) is further for higher kink initial velocities. This is because of the nature of the defect, whose effect is stronger for larger $\phi_t$. However, when both kinks approaches the defect from the loss side of the defect the final velocities of the kinks are remain unchanged.

\begin{figure*}[h!]
\begin{center}
  \centering
  \subfigure[Kink-defect collision from the gain side]{\includegraphics[width=0.45
 \textwidth]{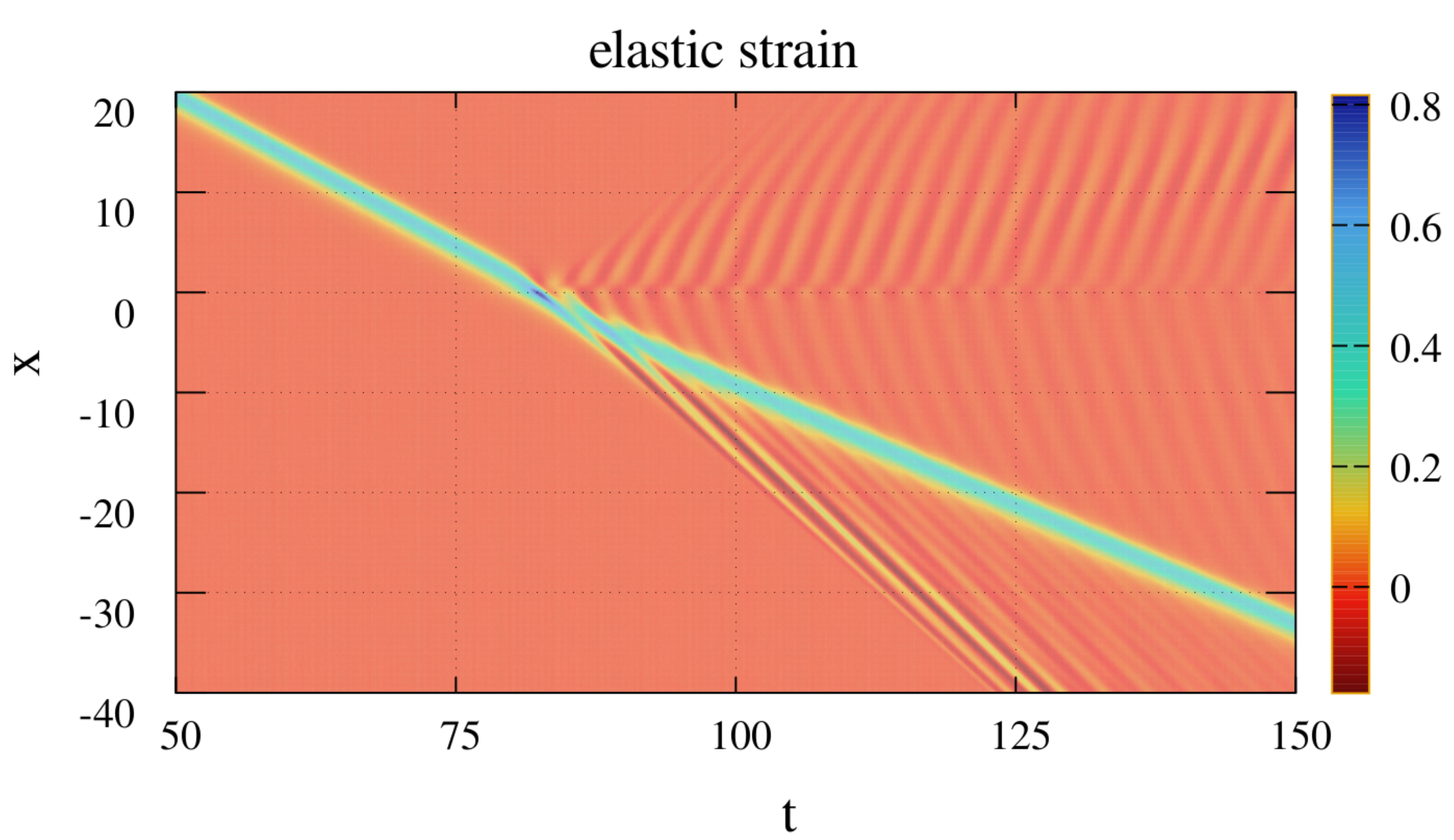}\label{fig:ESMAPKinkNat12000X500Vm060Beta2000Eps120}}
  \subfigure[Antikink-defect collision from the gain side]{\includegraphics[width=0.45
 \textwidth]{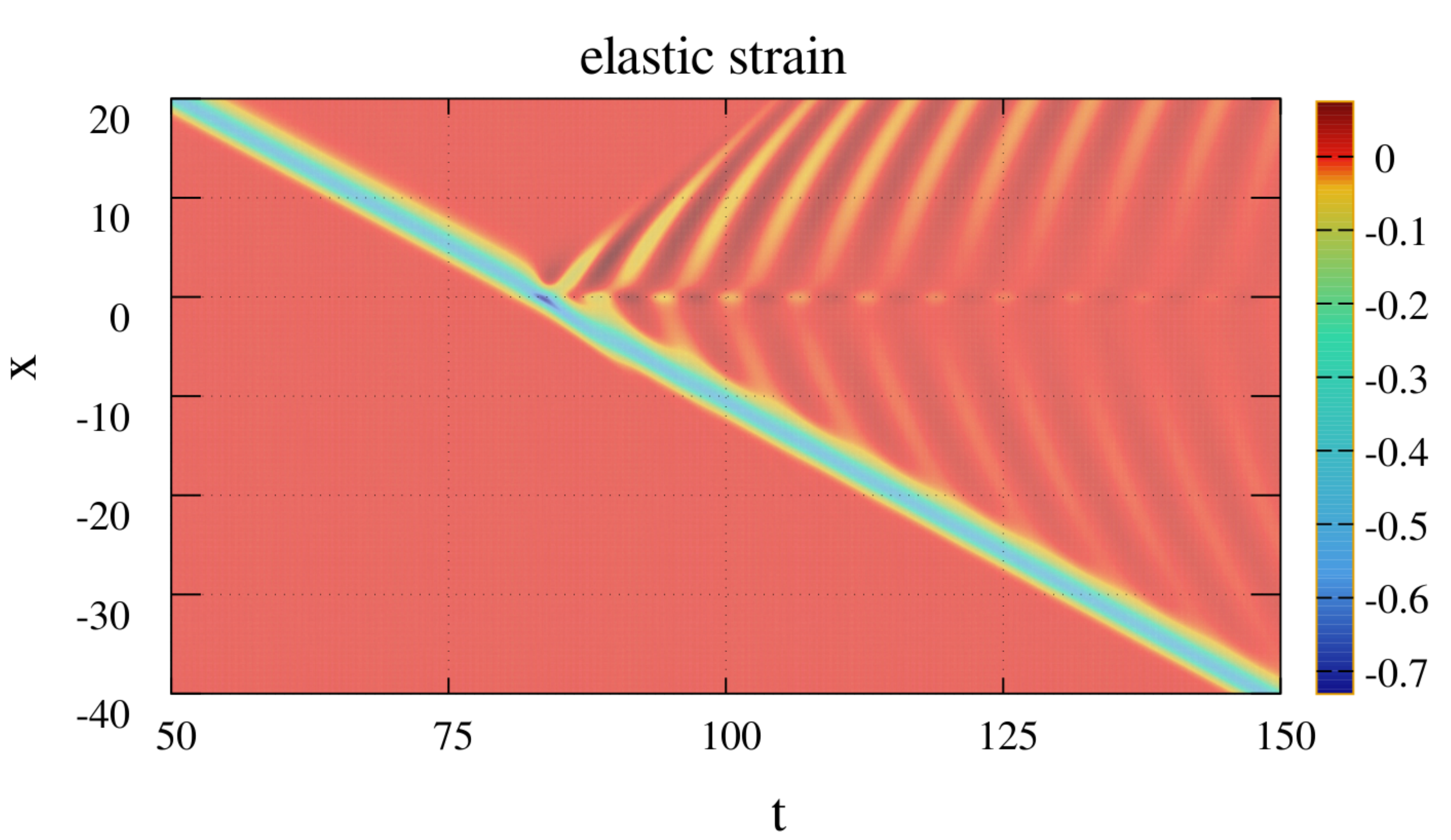}\label{fig:ESMAPantiKinkNat12000X500Vm060Beta2000Eps120}}
\\
 \subfigure[Kink-defect collision from the loss side]{\includegraphics[width=0.45
 \textwidth]{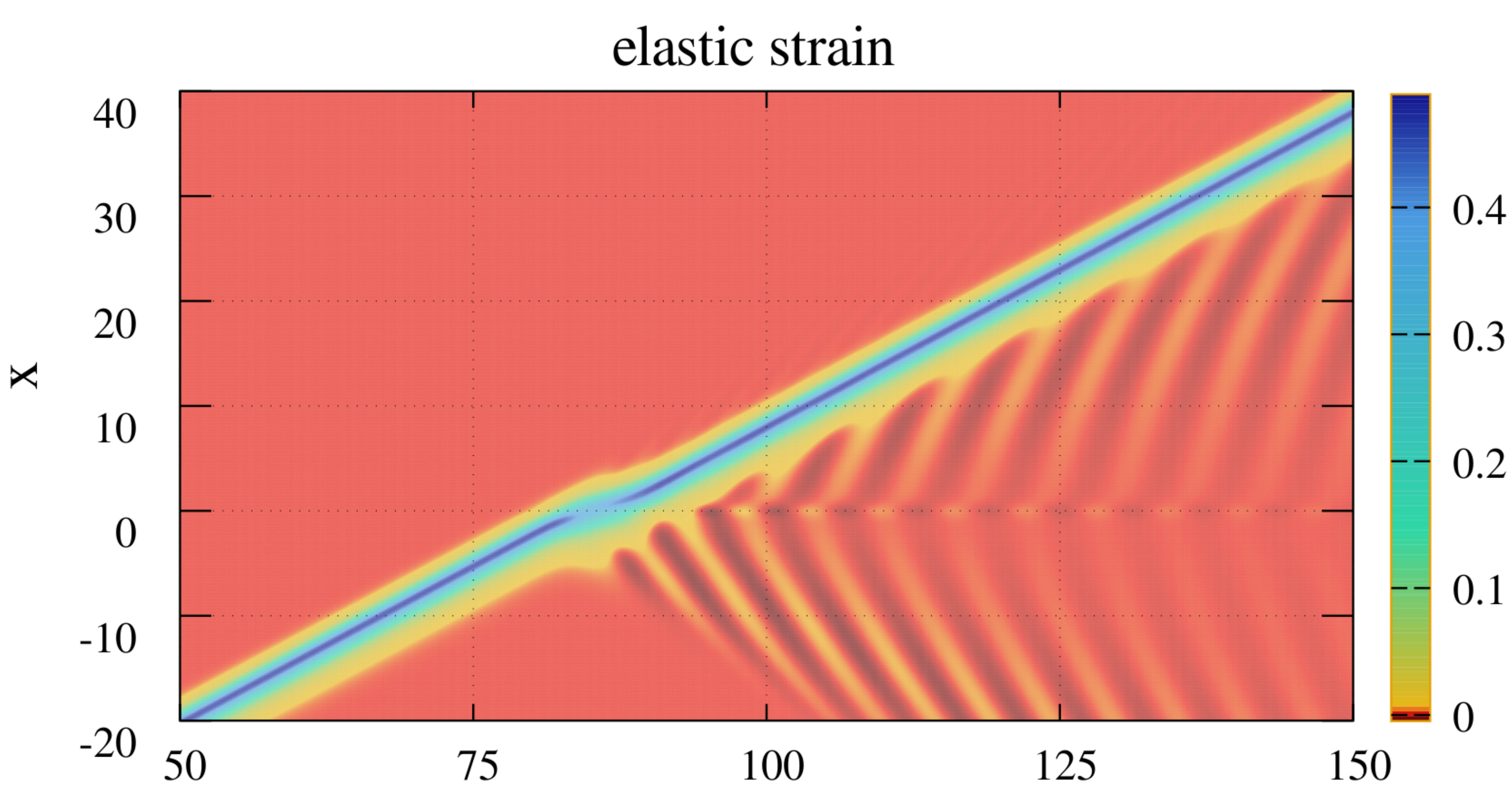}\label{fig:ESMAPKinkNat12000Xm500V060Beta2000Eps120}}
  \subfigure[Antikink-defect collision from the loss side]{\includegraphics[width=0.45
 \textwidth]{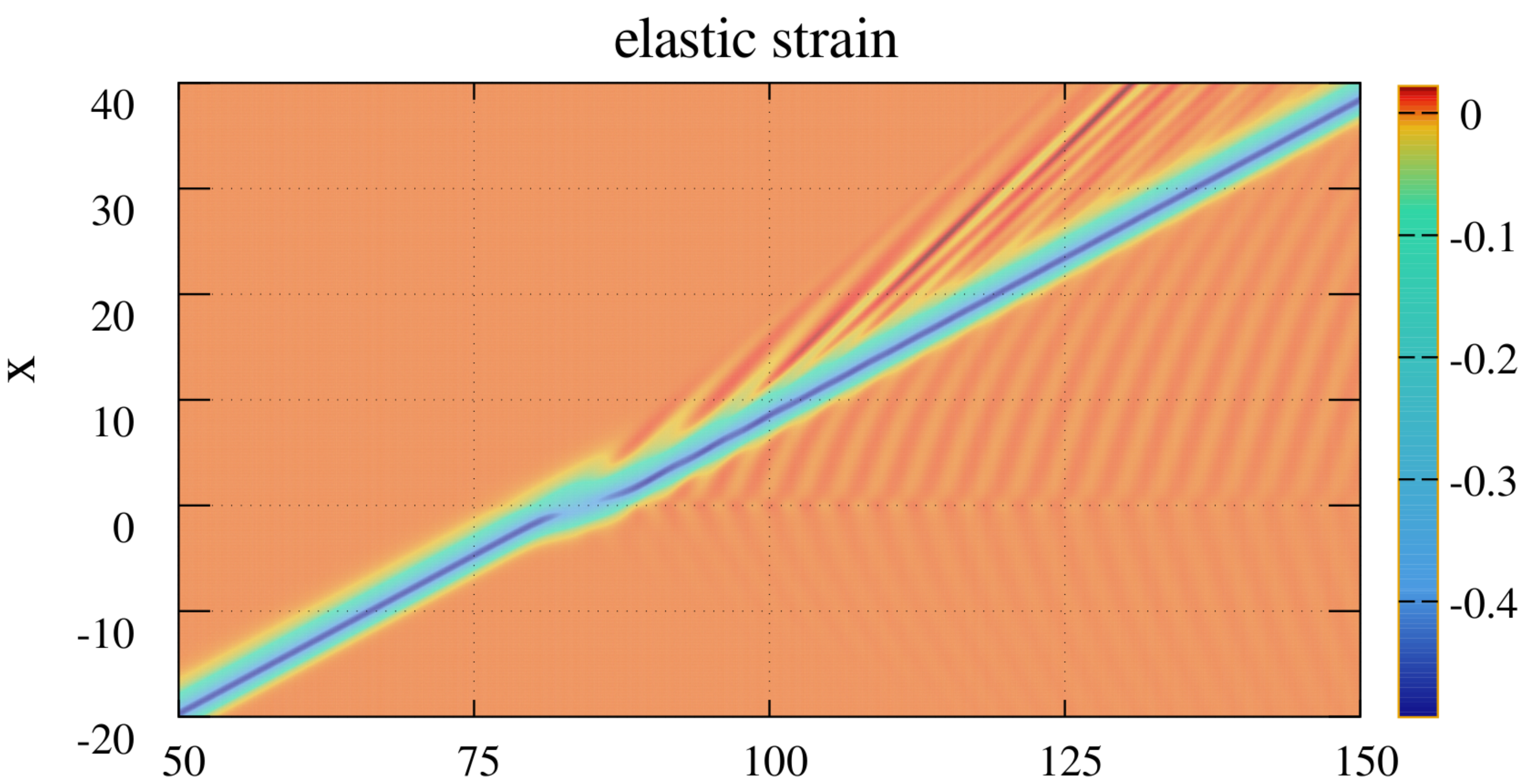}\label{ fig:ESMAPantiKinkNat12000Xm500V060Beta2000Eps120}}
\\
  \caption{Elastic strain $\phi_x$ for the kinks collision with defect. The initial conditions of the kinks are $x_0=\pm50$, $V=\pm0.6$. And for the defect parameters we used $\epsilon=1.2$ and $\beta=2.0$.  }\label{fig:ElasticStrain}
\end{center}
\end{figure*}

In order to confirm our findings, we plotted the space-time picture of the elastic strain $\phi_x$ in Fig.~\ref{fig:ElasticStrain} for the case when the kinks come from (a,b) the gain side (c,d) the loss side. In (a) elastic strain of the kink $\phi_{(0,1)}$ moves slower after the interaction with defect showing that the kink becomes wider after passing through the defect, in contrast to this, in (b) elastic strain of the kink $\phi_{(1,0)}$ moves faster and as a result the kink becomes narrower. From (c,d) it is clearly seen that the velocity of the kink elastic strain does not change after kinks pass through the defect and it is almost constant. It can be concluded that energy transmission between translational mode of the kink and its internal structure is possible when the kink moves toward the gain side of the defect. However, the results of numerical simulations demonstrate that this is not happen when the kink approaches the defect from the opposite direction.
\begin{figure}[h]
\centering
\includegraphics[width=1.0\textwidth]{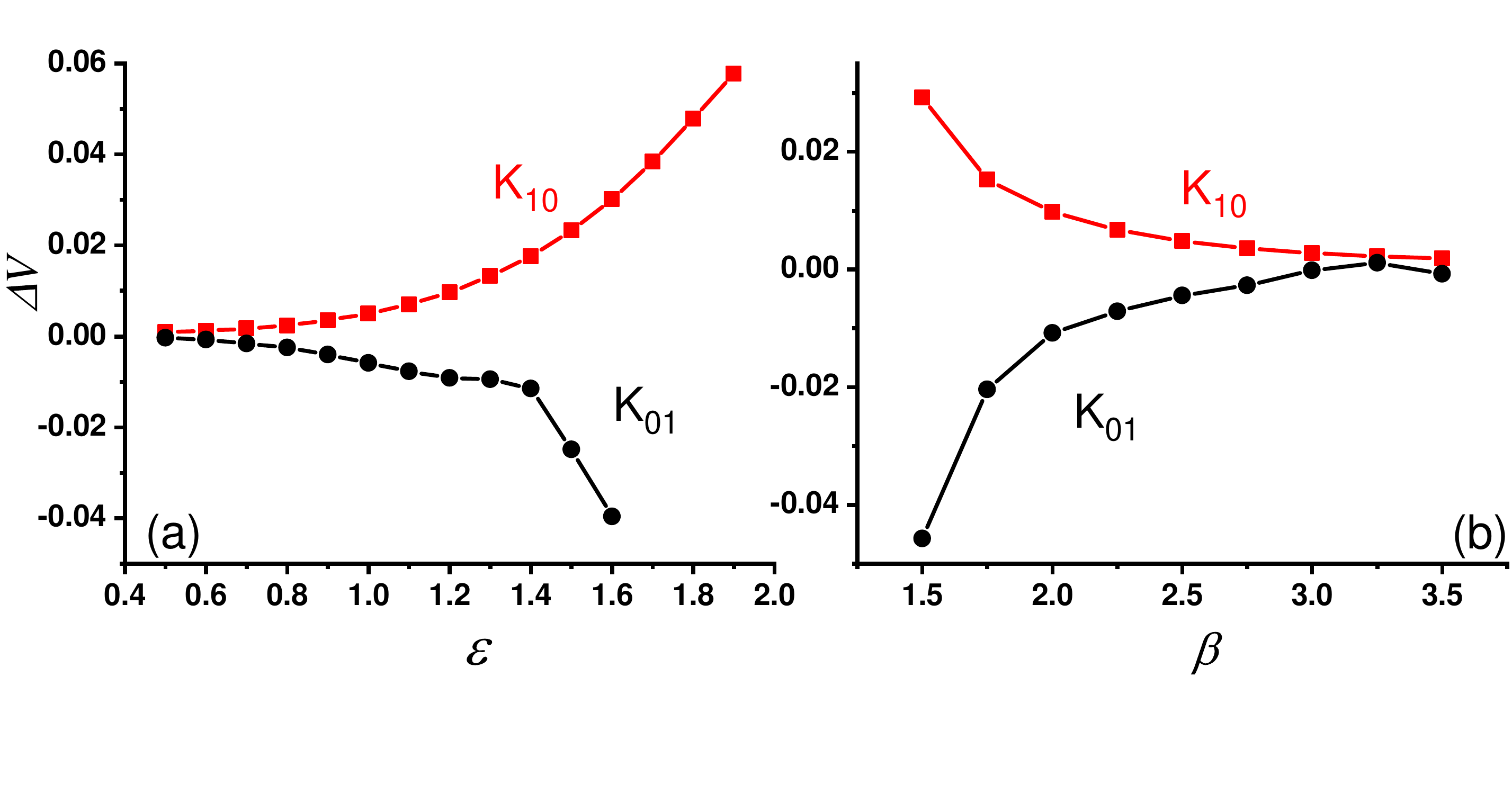} \caption{Kink velocity shift as the function of (a) defect amplitude $\epsilon$ and (b) defect inverse width $\beta$ for the case when kink comes from the gain side of the defect with initial velocity $V=0.4$. The perturbation parameters are $\beta=2.0$  in (a) and $\epsilon=1.2$ in (b).} 
\label{fig:VelocityShift}
\end{figure}
To explain the effect of the $\mathcal{PT}$-symmetric defect on the final velocity of the kinks, in Fig.~\eqref{fig:VelocityShift}, we have plotted the kink velocity shift $\Delta V=V_f-V_i$ as the function of defect parameters for the case when the kink comes from the gain side with initial velocity $V=0.4$. In (a) $\Delta V$ as the function of defect amplitude $\epsilon$ is plotted for the kinks $\phi_{(0,1)}$ and $\phi_{(1,0)}$. It can be seen that the kink velocity shift decreases (increases) with increasing defect amplitude $\epsilon$ for the kink $\phi_{(0,1)}$ ($\phi_{(1,0)}$). In (b) $\Delta V$ as the function of defect inverse width $\beta$ is plotted for the two kinks. The plot demonstrates that $\Delta V$ increases (decreases) with increasing defect inverse width $\beta$ for the kink $\phi_{(0,1)}$ ($\phi_{(1,0)}$). 

\begin{figure}[h]
\centering
\includegraphics[width=0.8\textwidth]{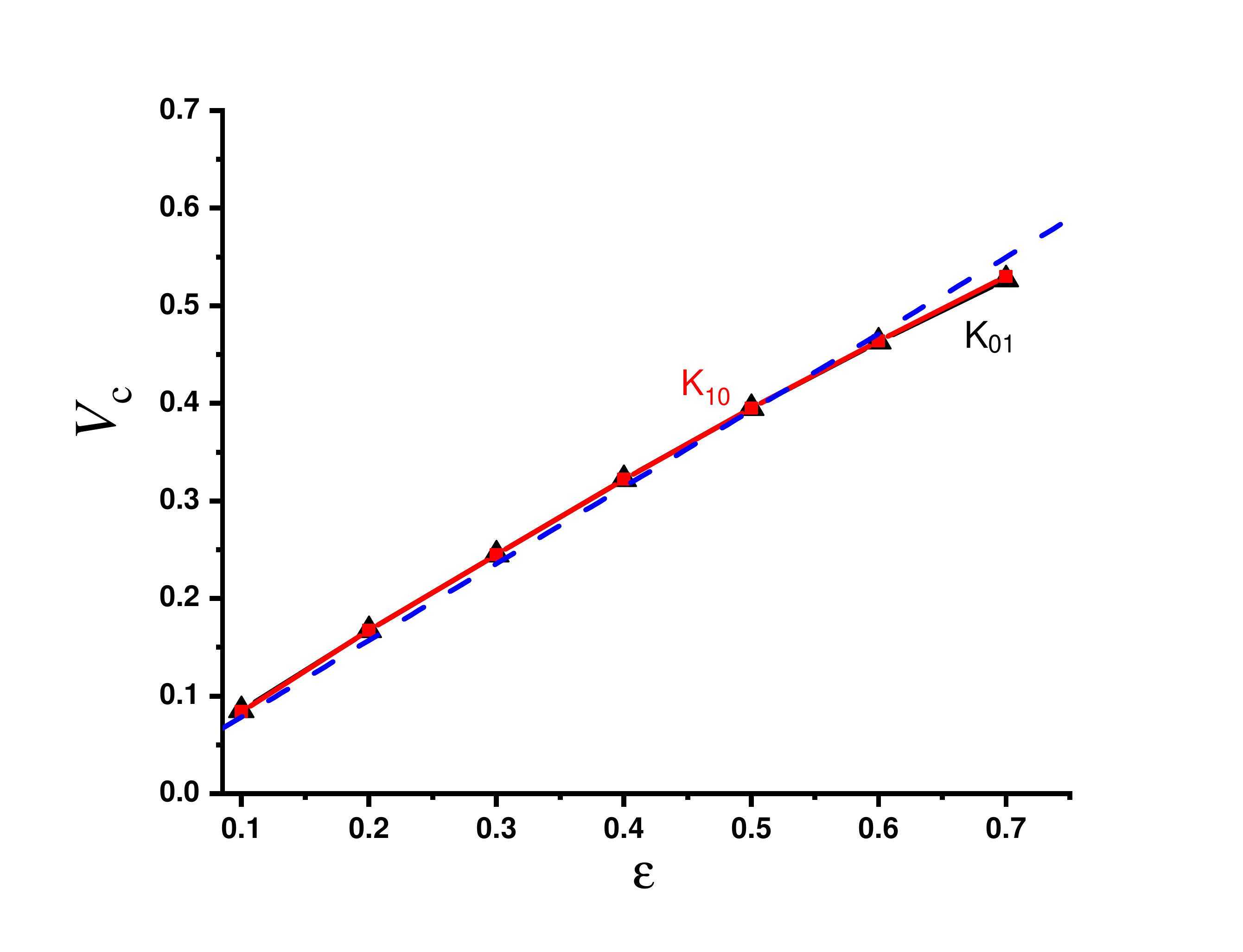} 
\caption{The critical velocity of the kink as function of the defect amplitude for the kink comes from the initial position of $x_0=-15$. The perturbation parameter is $\beta=1.0$. Solid line is for the continuum system, while dashed line is for the collective variable
method.} 
\label{fig:CriticalVelocity}
\end{figure}
A kink approaching the defect from the gain side always passes, while in the opposite direction it must have sufficiently large initial momentum to pass through the defect instead of being trapped in the loss region. Therefore, for the kink coming from the loss side there must have been a critical velocity above which kink passes through the defect and below which it can be trapped. In Fig.~\ref{fig:CriticalVelocity} the plane of the parameters $\epsilon$ and $V_c$ is shown for the two kinks belong to different sectors (0,1) and (1,0). The chart separates the two possible scenarios of the kink-defect interaction when the kink approaches the defect from the lossy side. Above the chart kink has a sufficient initial momentum and it passes through it. Contrary to this, below the chart the kink always trapped in the lossy region of the defect and eventually stops. The dash line demonstrates the results obtained with the help of collective variable method Eq.~\eqref{CriticalV} is in perfect agreement with the result of the continuum system.

\begin{figure*}[h!]
\begin{center}
  \centering
  \subfigure[$V=-0.1500$]{\includegraphics[width=0.45
 \textwidth]{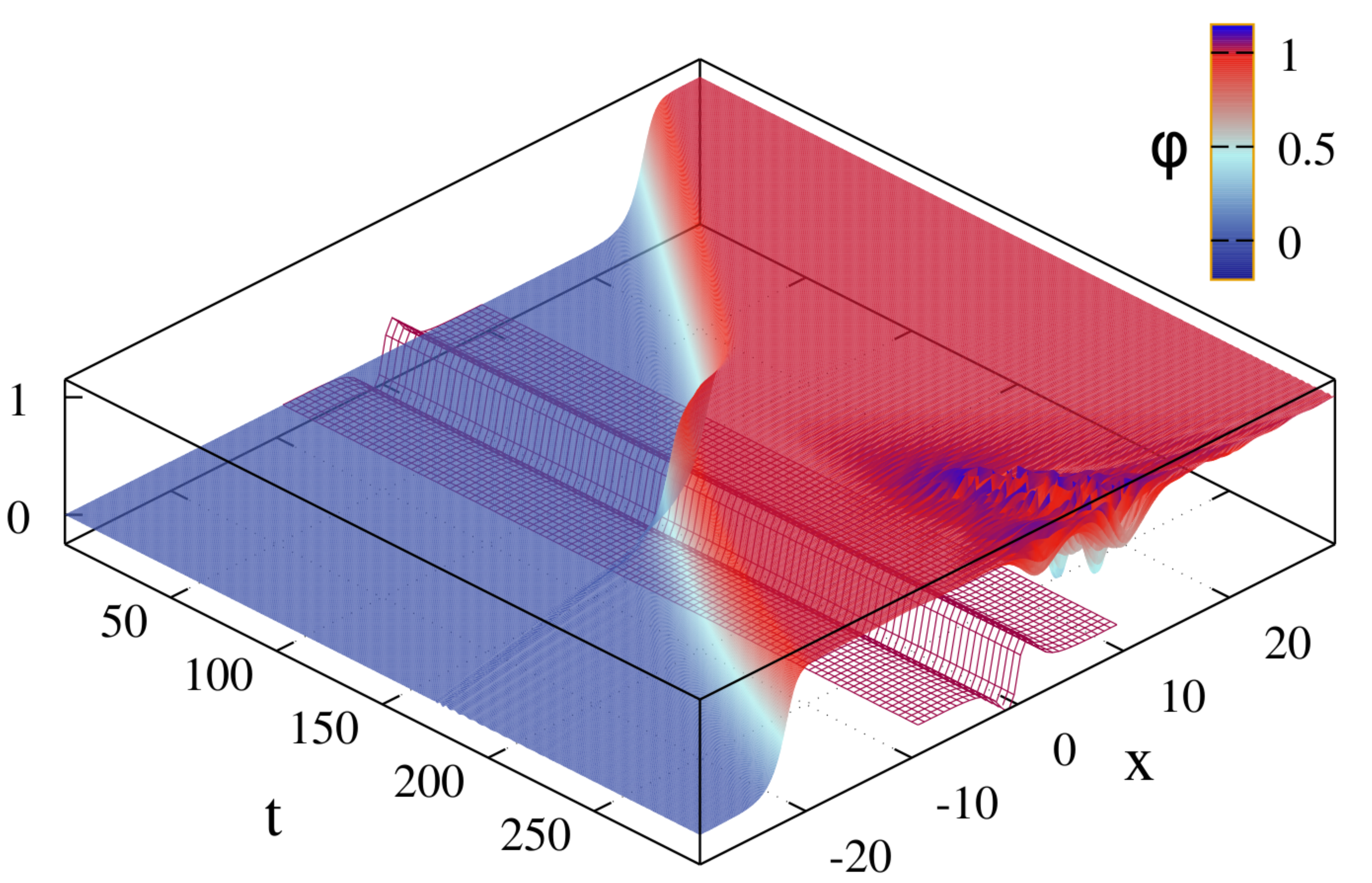}\label{fig:KinkNat5000Xm250Vm015Beta0750Eps05}}
\\
  \subfigure[$V=-0.2870$]{\includegraphics[width=0.45
 \textwidth]{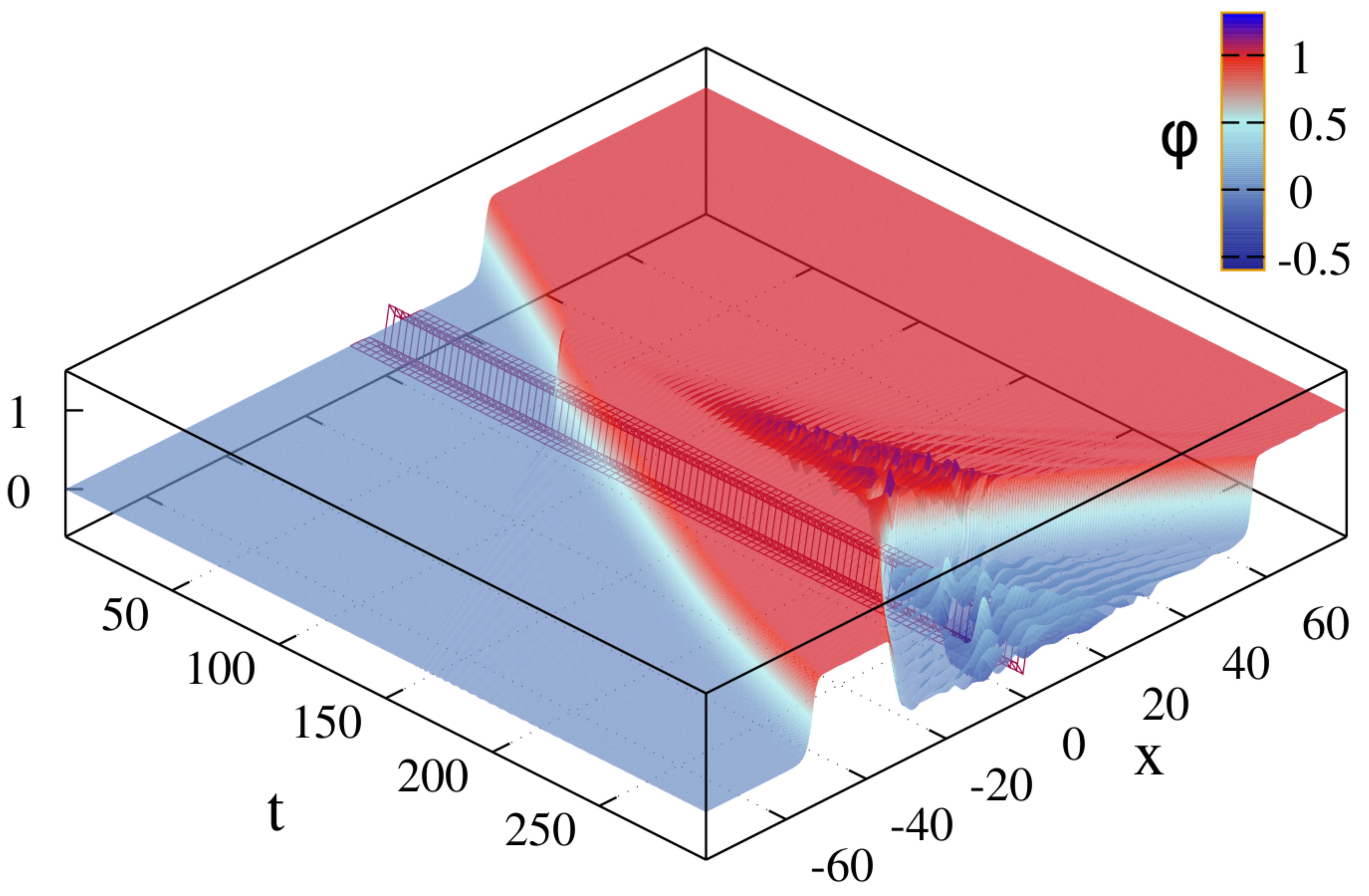}\label{fig:KinkNat5000Xm250Vm02870Beta0750Eps05}}
 \subfigure[$V=-0.2875$]{\includegraphics[width=0.45
 \textwidth]{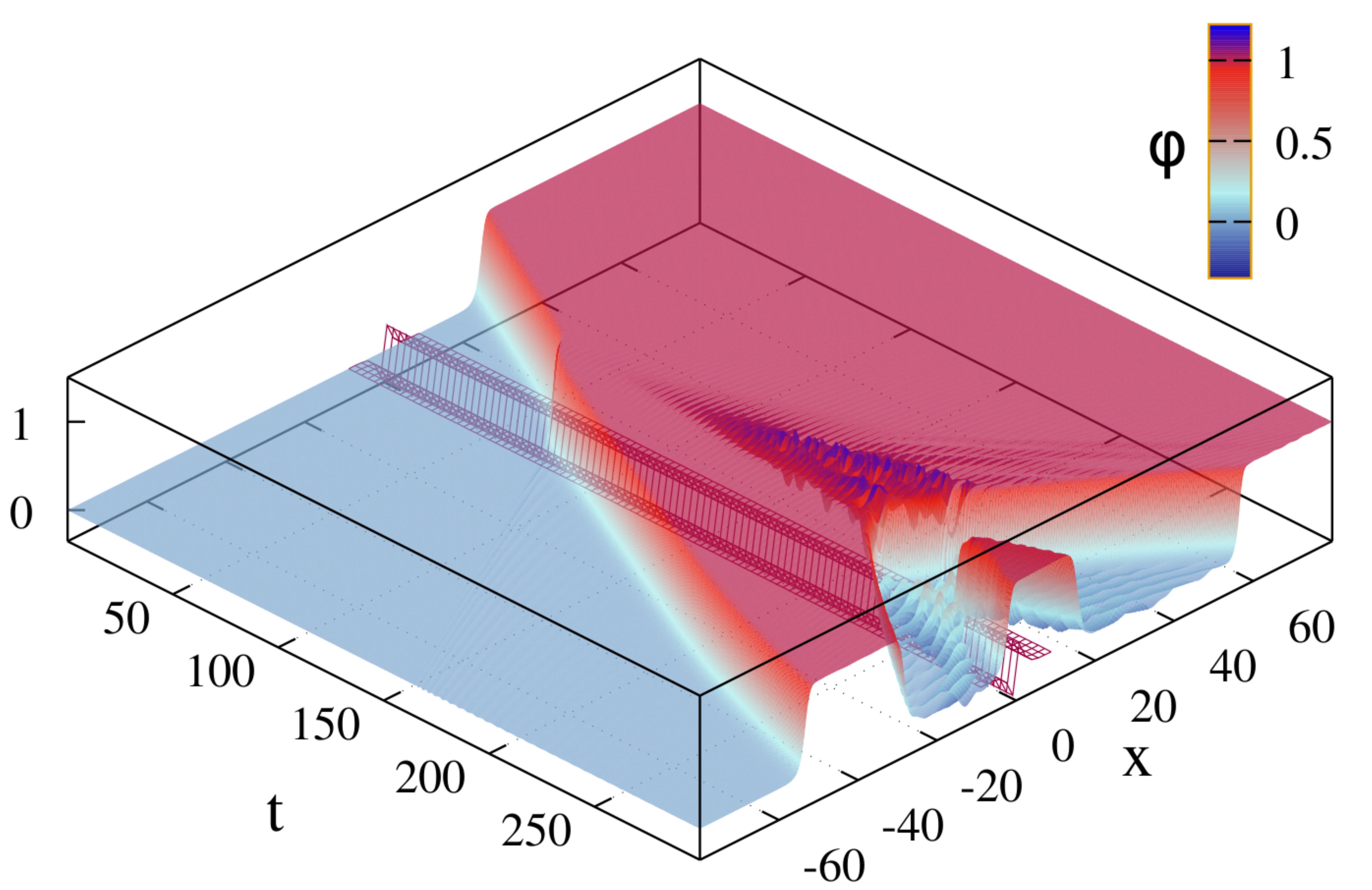}\label{fig:KinkNat5000Xm250Vm02875Beta0750Eps05}}
\\
  \caption{Space-time picture of the kink interaction with a $\mathcal{PT}$-symmetric for various kink initial velocities, as indicated for each curve. The parameters of the defect are $\epsilon=0.5$ and $\beta=0.75$  }
  \label{fig:3Dplot}
\end{center}
\end{figure*}
In Fig.~\ref{fig:3Dplot} the space-time pictures of interaction of the kink $\phi_{(0,1)}$ with a $\mathcal{PT}$-symmetric defect with amplitude  $\epsilon=0.5$ are shown for the case when kink comes from the gain side of the defect. The kink starts from the initial position $x_{0}=25$ and moves toward the defect with initial velocities (a) $V=-0.15$, (b) $V=-0.2870$ and (c) $V=-0.2875$. In (a) kink interacts with defect and after passing through it, small amplitude waves (phonons) are created inside the gain side of the defect for a short time and as a result they are amplified and eventually scattered during the whole chain. In (b) phonons live for a longer time in the gain side of the defect and they acquire energy from the defect and at time $t\approx205$ kink-antikink pair are formed in the final state. At larger values of $V$ in (c) two pairs of kink-antikink with interval time of $t\approx25$ are formed in the final states.

\begin{figure*}[h!]
\begin{center}
  \centering
  \subfigure[Kink $\phi_{01}$]{\includegraphics[width=0.49
 \textwidth]{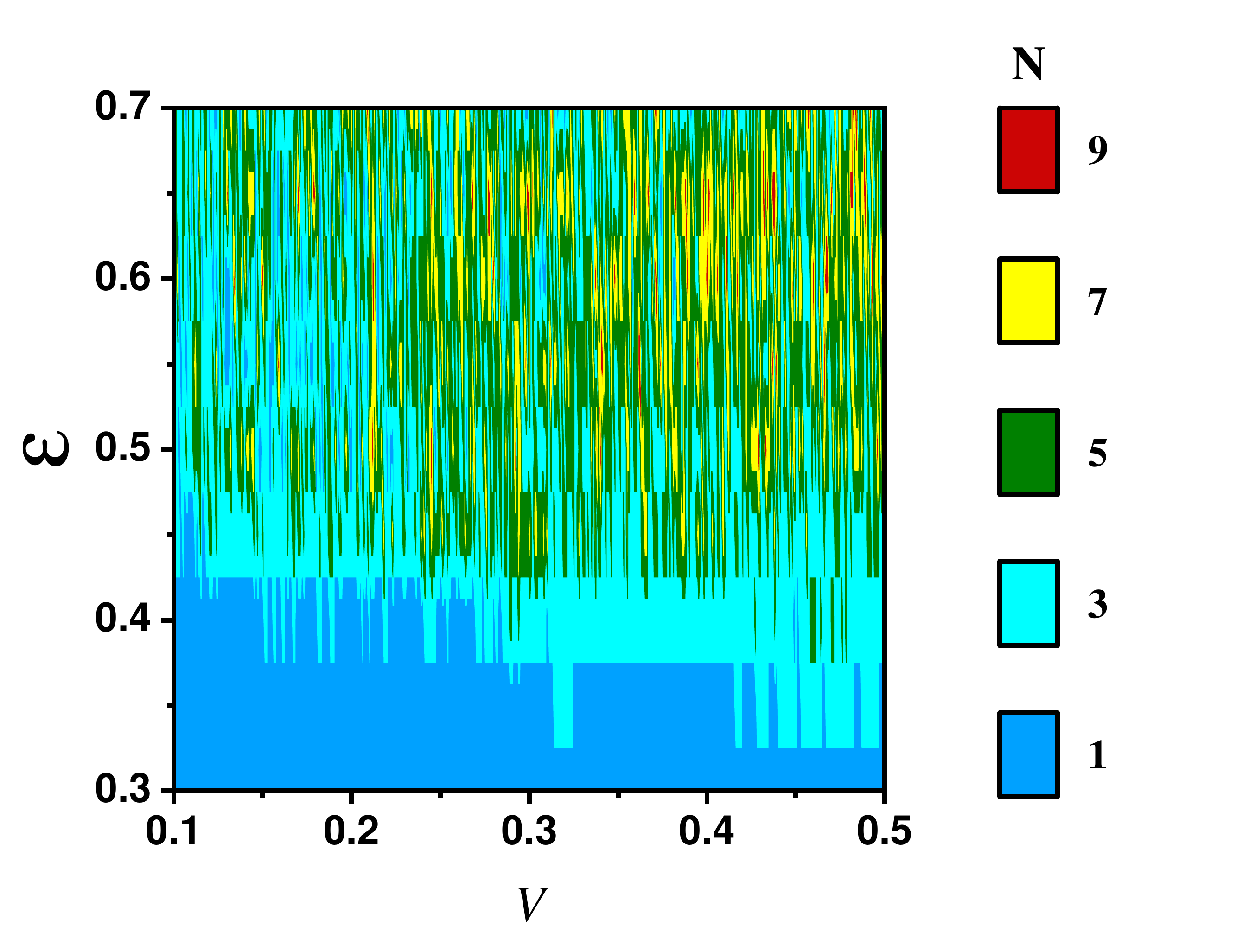}\label{fig:NumberofKink01}}
  \subfigure[Antikink $\phi_{10}$]{\includegraphics[width=0.49
 \textwidth]{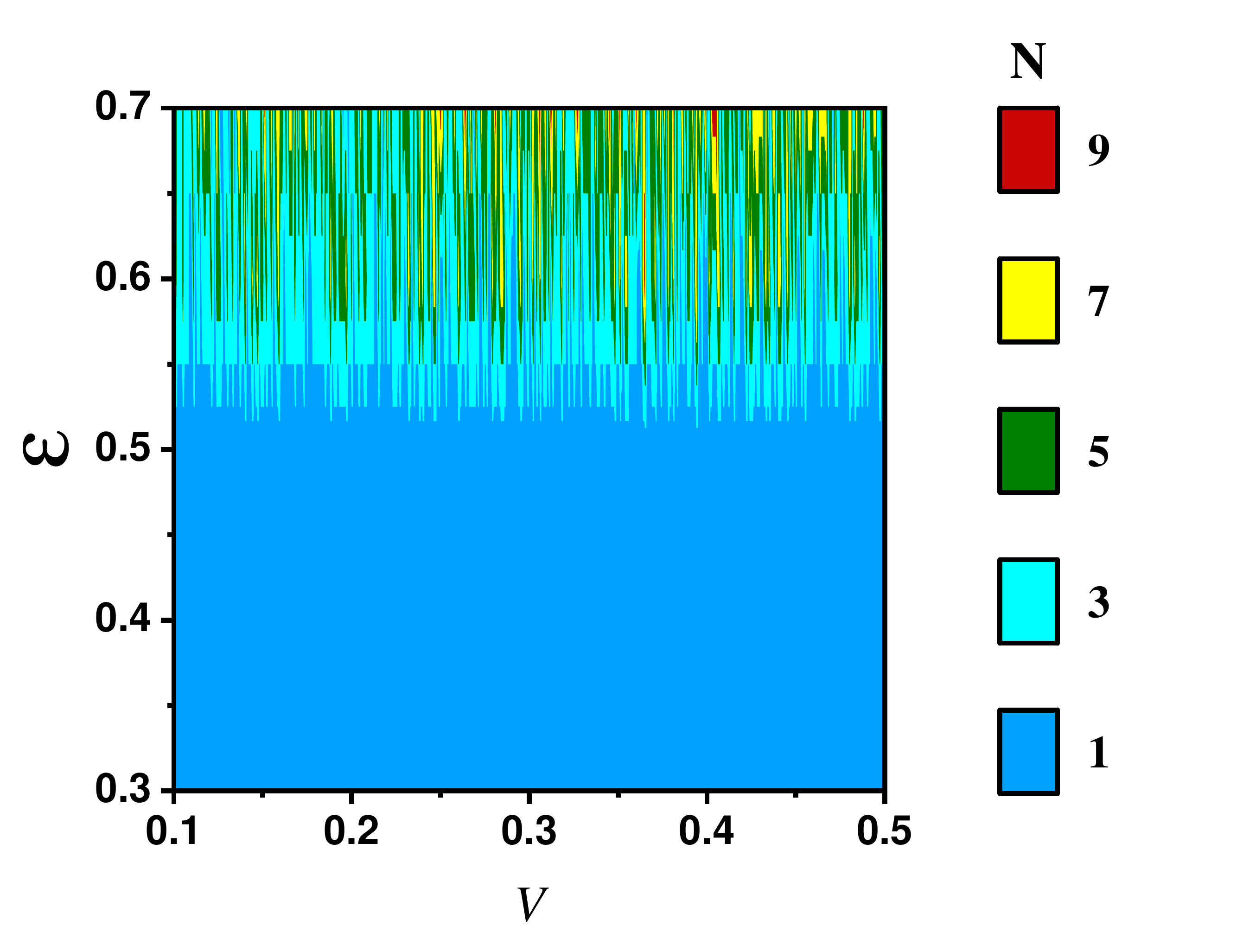}\label{fig:NumberofKink10}}
\\
  \caption{Contour plot for (a) kink and (b) antikink interaction with a $\mathcal{PT}$-symmetric defect for the case when kink comes from the gain side. The parameter of defect is $\beta=0.75$ in this case.}
  \label{fig:NOK}
\end{center}
\end{figure*}

Notice that numbers of the kink-antikink pairs created in the final state of the interaction clearly depend on the amplitude of the defect $\epsilon$ and also the initial velocity of the kink $V$. To illustrate this point, we have plotted the number of kinks $N$ produced after the interaction with defect as the function of kink initial velocity and amplitude of the defect in the contour plot of Fig.~\ref{fig:NOK}. We note that the creation of the kinks with the same topological charge is forbidden due to topological constraint, hence, kinks can be created in kink-antikink pairs only. The graph shows the results for the case when (a) kink and (b) antikink come from the gain side of the defect and the numerical simulations are run until the time $t=400$. It is clearly seen that the number of kinks in the final state increases with increasing $\epsilon$ and $V$. However, for an antikink $\phi_{(1,0)}$ this increase occurs for higher amplitudes, i.e., $\epsilon > 0.5$; see Fig.~\ref{fig:NOK}~(b). This means that the formation of kink-antikink pairs from phonons happens more hardly for interaction of antikink with defect. 

We have performed the same numerical simulations for different values of initial velocities of the kink for the case when kink comes from the loss side of the defect. Our findings demonstrate that the creation of the kink-antikink pairs in the final state of the interaction are not formed at all; at least for small values of defect amplitudes. This is due to the fact that when the kink hits the defect from the this side phonons are trapped by loss region and finally they are destroyed.

\section{Kink-anitkink interaction with defect}\label{Sec:Kink-anitkink}
The next step it to study the interaction of the kink $\phi_{(0,1)}$ and antikink $\phi_{(1,0)}$ with a $\mathcal{PT}$-symmetric perturbation. Here, we use the configuration (1,0,1) which composed of one kink comes from the gain side and one antikink comes from the loss side of the defect. For the configuration (0,1,0) the positions of the kink and antikink are reversed (see \cite{Moradi.JHEP.2017}). The kinks are initially placed at $x=x_1$ and $x=-x_2$ and moving toward each other with initial velocities $V=-V_i$ and $V=V_i$, respectively. Since there is no exact two soliton solution in the $\phi^6$ model, we use the following initial configuration for the sector (1,0,1)
\begin{equation}\label{Kink101}
\phi_{(1,0,1)}(x,t) = \phi_{(1,0)}\left(\frac{x+x_1-V_it}{\sqrt{1-V_i^2}}\right)+\phi_{(0,1)}\left(\frac{x-x_2+V_it}{\sqrt{1-V_i^2}}\right).
\end{equation}
For the sector (0,1,0), we have
\begin{equation}\label{Kink010}
\phi_{(0,1,0)}(x,t) = \phi_{(0,1)}\left(\frac{x+x_1-V_it}{\sqrt{1-V_i^2}}\right)+\phi_{(1,0)}\left(\frac{x-x_2+V_it}{\sqrt{1-V_i^2}}\right)-1.
\end{equation}
The initial separation distance between kink and antikink is equal to $x_1+x_2$. The above configurations are not exact solutions of the equation of motion \eqref{eq:eom}, however, for $x_1+x_2\gg ~1$ the overlap between the kinks is exponentially small, therefore Eqs.~\eqref{Kink101} and \eqref{Kink010} are exact solutions up to exponentially small terms. 

It is well known that there exist two different
scattering channels in the scattering problem: (i) bion formation, where kink and antikink become trapped after the collision for initial velocities less than $V_{cr}$ (ii) kink reflection, where kink and antikink collide with initial velocity larger than $V_{cr}$ and finally recede to infinity after the collision.

\begin{figure*}[h!]
\begin{center}
  \centering
  \subfigure[$x_0=-15$, $V=0.25$ for kink and $x_0=20.0$, $V=-0.25$ for antikink.]{\includegraphics[width=0.45
 \textwidth]{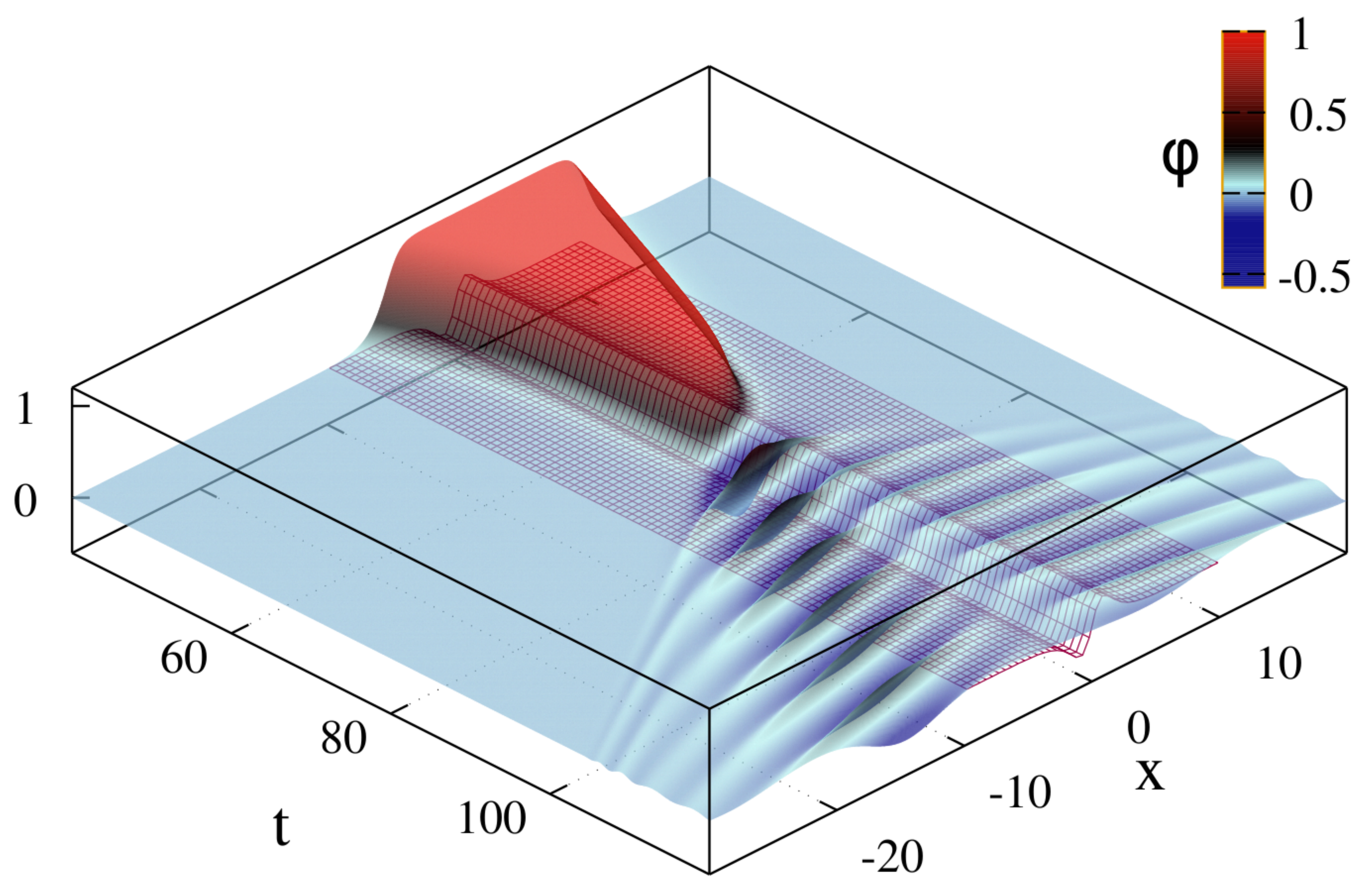}\label{fig:KinkXm150V025AntikinkX200Vm025Beta1Eps055}}
  \subfigure[$x_0=-15$, $V=0.25$ for antikink and $x_0=20.0$, $V=-0.25$ for kink.]{\includegraphics[width=0.45
 \textwidth]{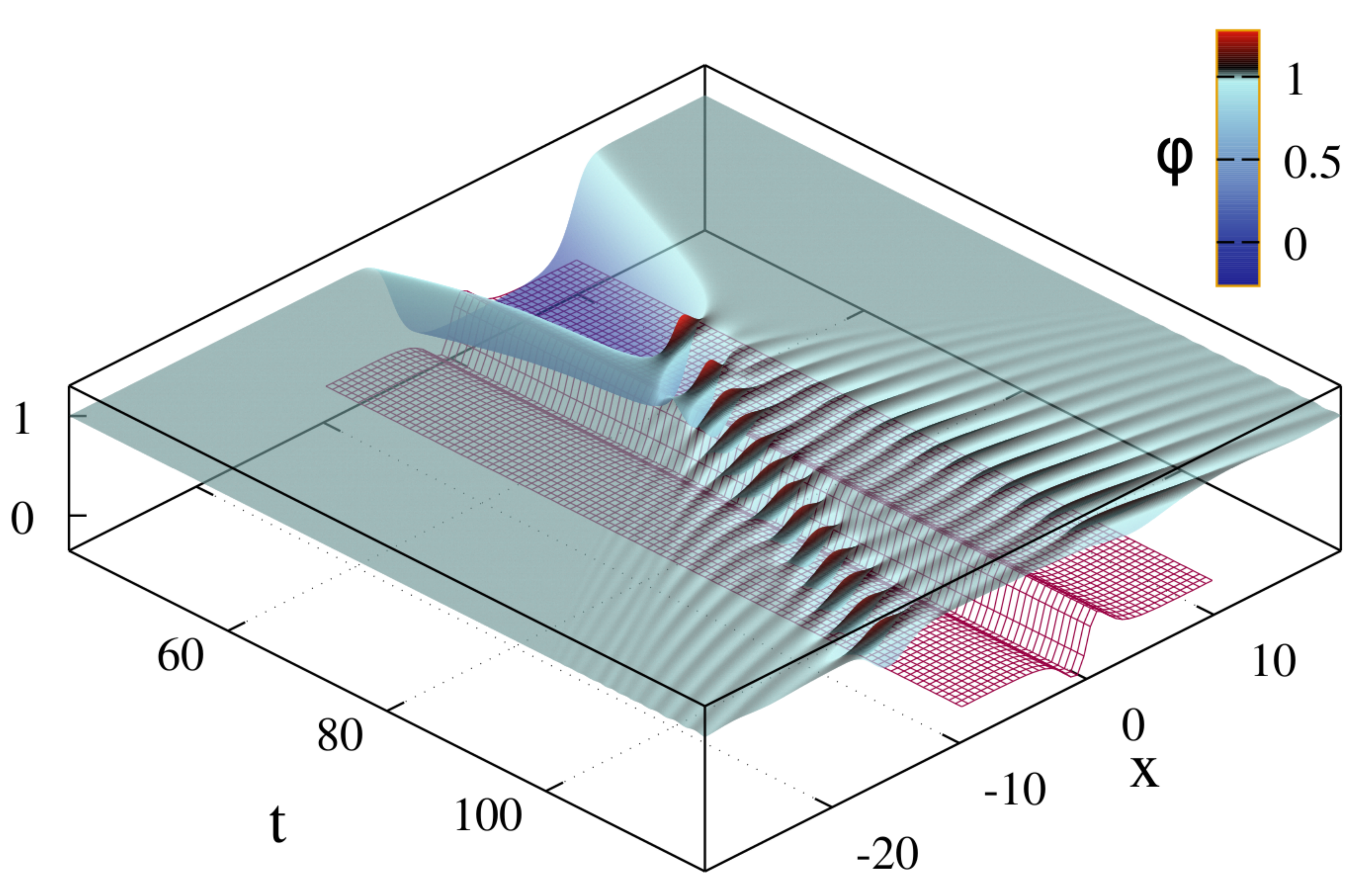}\label{fig:AntiKinkXm150V025kinkX200Vm025Beta1Eps055}}
\\
\subfigure[$x_0=-15$, $V=0.50$ for kink and $x_0=20.8$, $V=-0.50$ for antikink.]
{\includegraphics[width=0.45\textwidth]{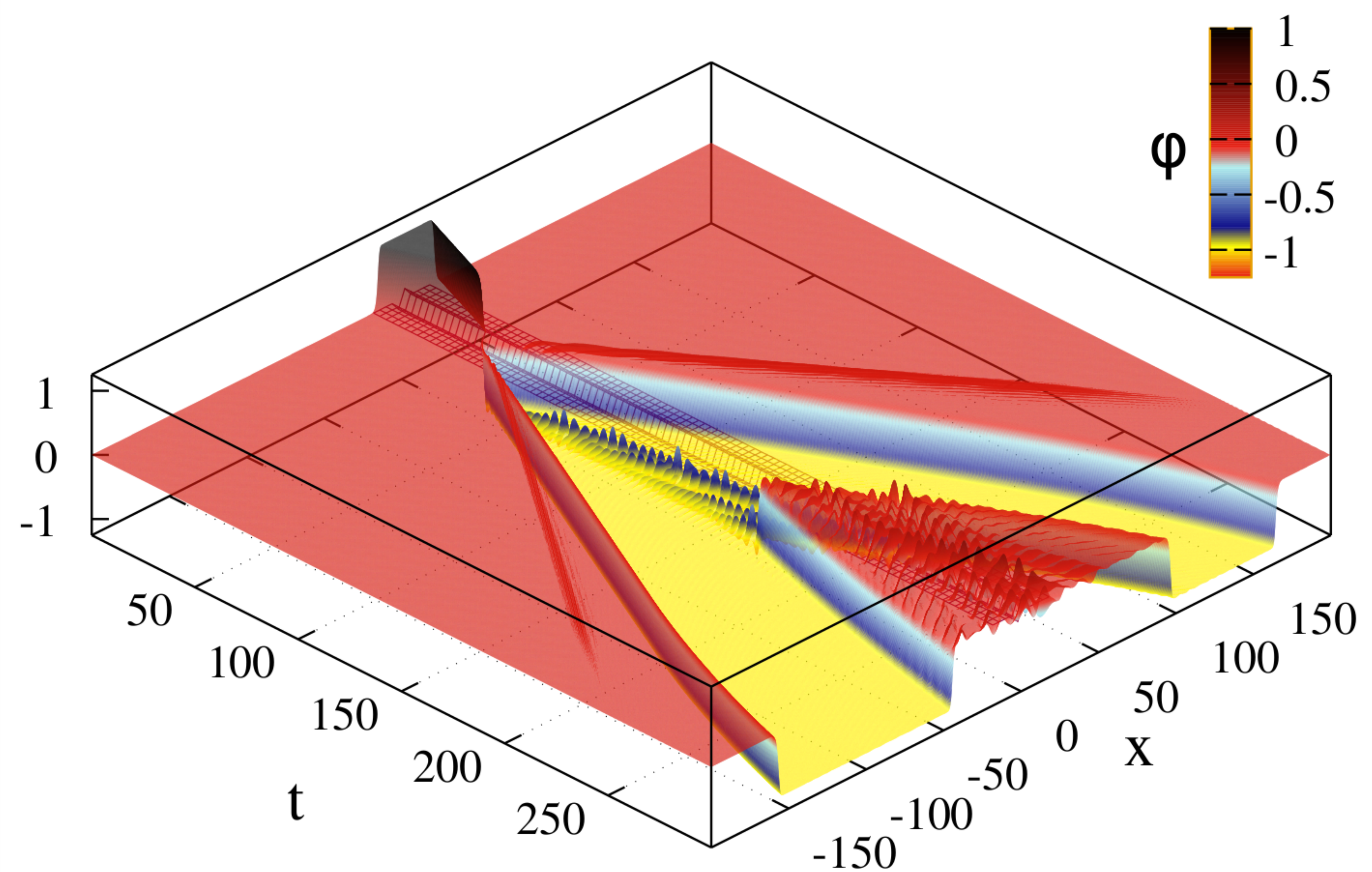}\label{fig:KinkXm150V050AntikinkX208Vm050Beta1Eps055}}
  \subfigure[$x_0=-15$, $V=0.50$ for antikink and $x_0=20.8$, $V=-0.50$ for kink.]
{\includegraphics[width=0.45\textwidth]{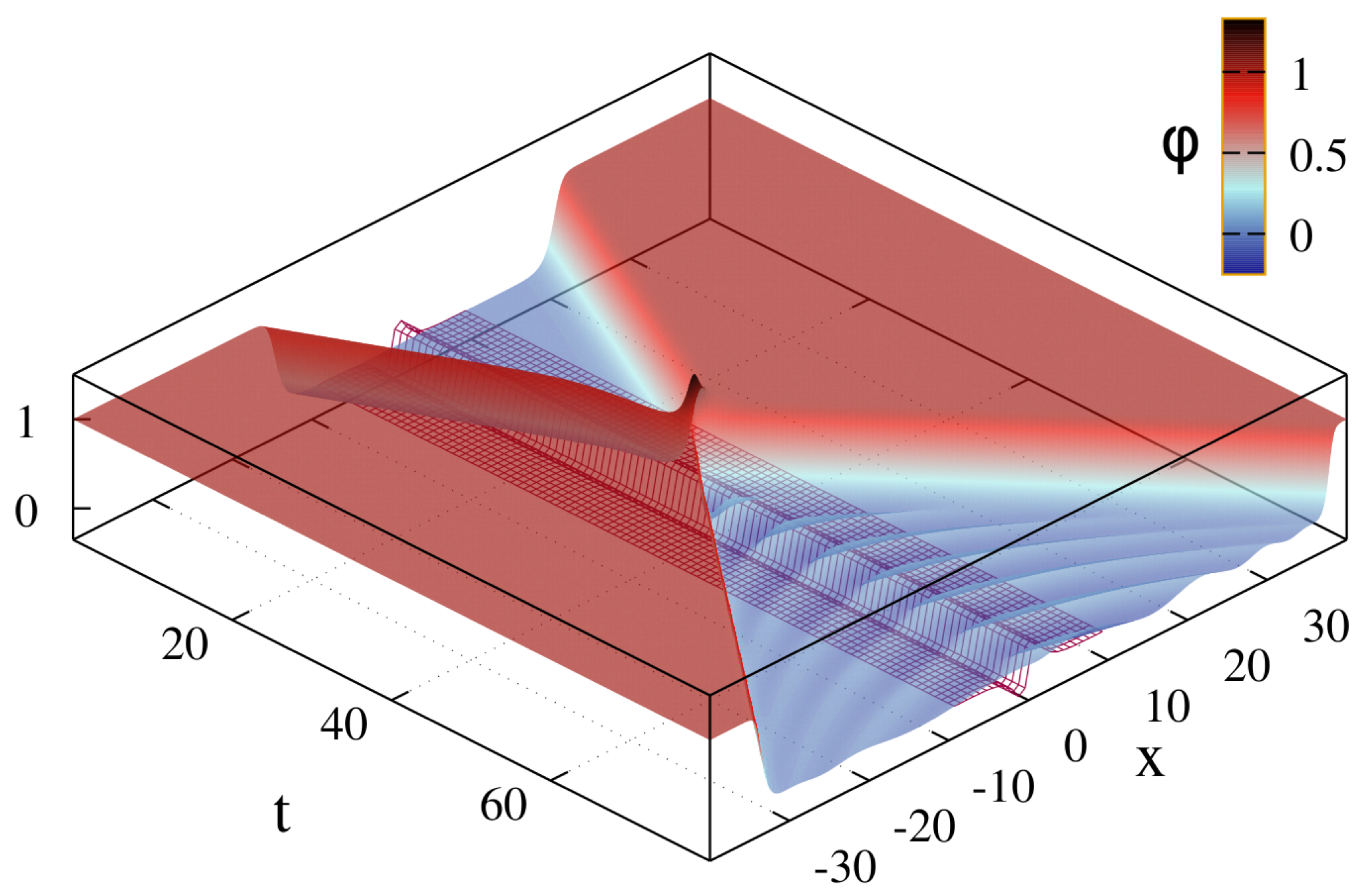}\label{fig:AntiKinkXm150V050KinkX208Vm050Beta1Eps055}}
  \\
  \subfigure[$x_0=-15$, $V=0.50$ for kink and $x_0=20.81$, $V=-0.50$ for antikink.]
{\includegraphics[width=0.45\textwidth]{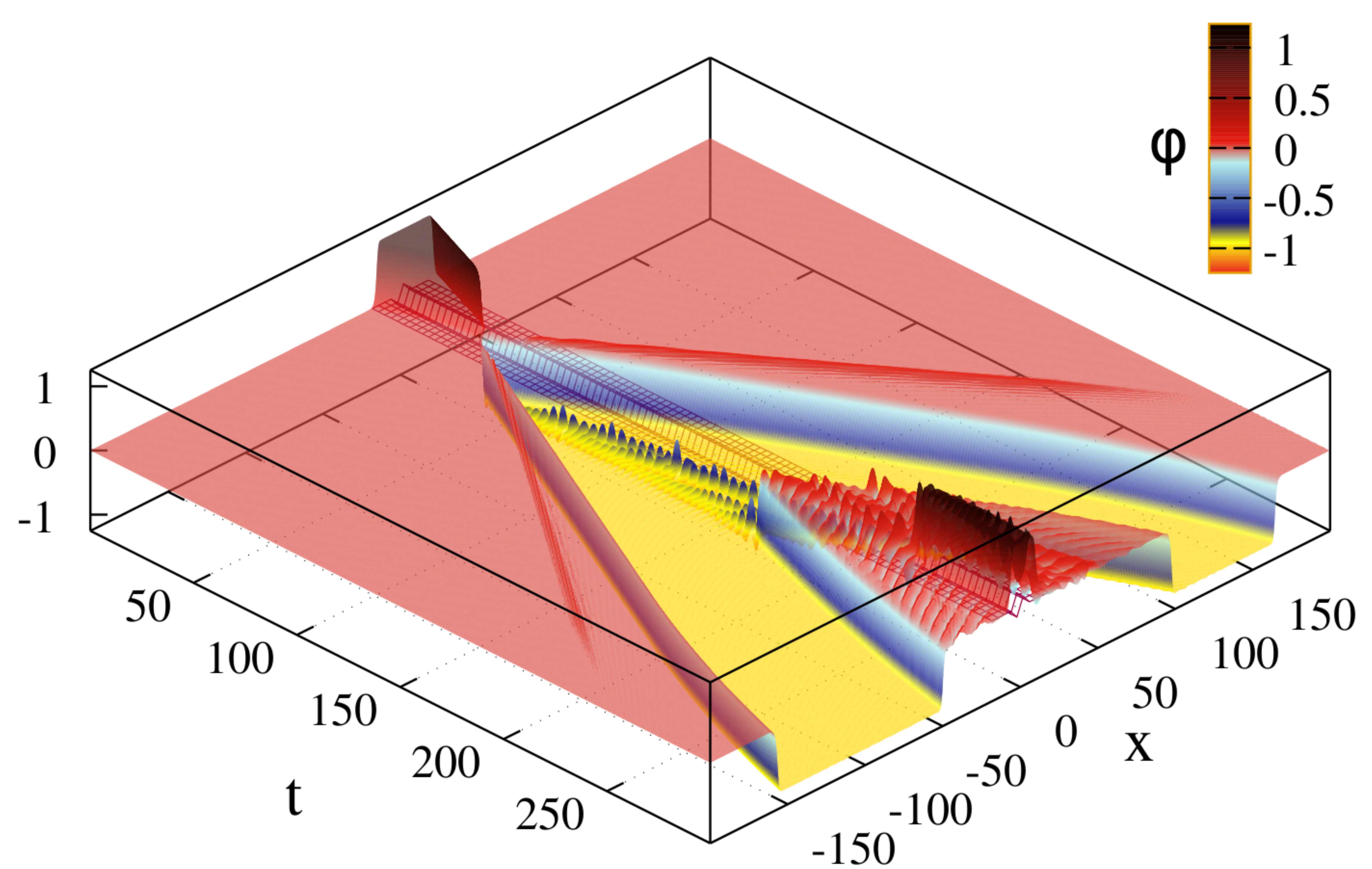}\label{fig:Xm15V05X2081Vm05Beta1Epcilon055}}
  \subfigure[$x_0=-15$, $V=0.50$ for antikink and $x_0=20.81$, $V=-0.50$ for kink.]
{\includegraphics[width=0.45\textwidth]{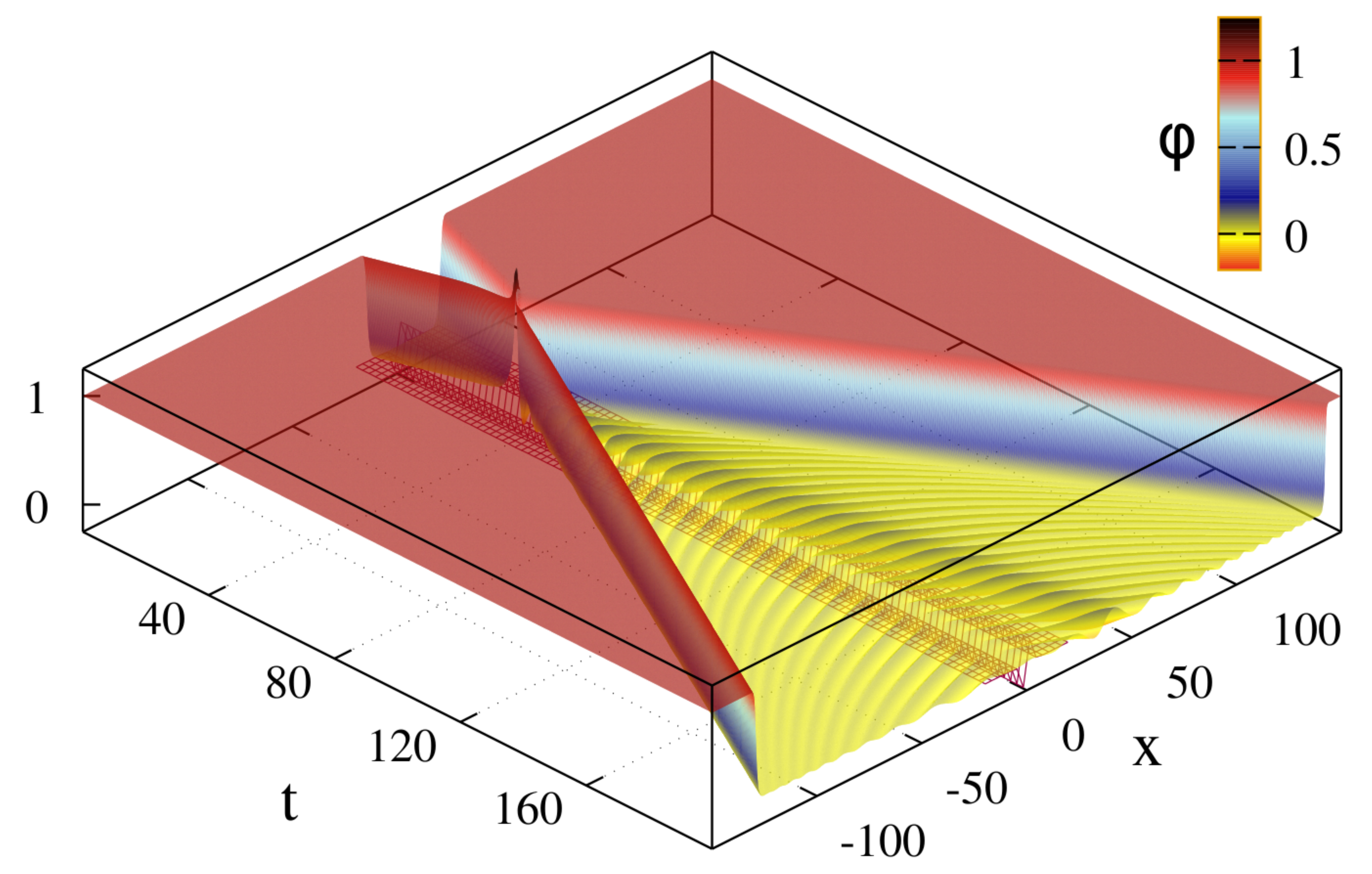}\label{fig:Xm15V05X2081Vm05Beta1Epcilon055-akk}}
  \\
  \caption{Space-time picture of kink and antikink interaction with a $\mathcal{PT}$-symmetric defect. The perturbation parameters are $\epsilon=0.55$ and $\beta=1$. The initial velocities and initial positions of the kinks are indicated for each plots.}
  \label{fig:2kcollision}
\end{center}
\end{figure*}

The numerical results of the kink-antikink collision for configuration (0,1,0) are presented in Figs.~\ref{fig:2kcollision} (a,c,e) and we give the results for the configuration (1,0,1) in Figs.~\ref{fig:2kcollision} (b,d,f). Two kinks approach the defect with amplitude $\epsilon=0.55$ from the both side and the critical velocity for which kink passes through the loss region can be easily found from Eq.~\eqref{CriticalV}, $V_c=0.4319$. The initial velocities and initial positions of each kinks are shown in the caption and we have performed numerical simulations during the scattering processes until time $t=400$. 

In (a,b) kinks have velocity $V=0.25$ which is below the threshold value, so the first kink (which comes from loss side) is trapped and the second one passes through the gain side of the defect and finally the collision takes place in the loss region, in consequence only phonons are figured out in the collision outcomes. In (c-f) the collision events happens in the gain part because kink and antikink have a enough initial momentum to pass through the loss part. This lead to the interesting phenomenons, which, to the best of our
knowledge, have not been reported for the kink-defect interaction systems. When kink and antikink collide in the gain region, first kinks escape from each other after a collision, then at time $t\approx160$ the second pair of the kink-antikink is formed; see Fig.~\ref{fig:2kcollision}~(c). The outputs obtained in Fig.~\ref{fig:2kcollision}~(e) indicate that in addition to the second pair of the kink-antikink, formation of a localized bound state, bion, is observed at time $t\approx220$. Since kink and antikink are mutually attractive quasi-particles, the second pair of kinks which is created from phonons affects the dynamic of the first pair of kinks. This is clearly seen in Figs.~\ref{fig:2kcollision}~(c,e) in which the velocity of kinks is not constant after the interaction with defect. In Figs.~\ref{fig:2kcollision}~(d,f) kink and the antikink collide at the gain region, recede from each other and escaping from the collision point with higher velocities $V_{1f}=-0.8$ and $V_{2f}=0.87$.

\section{Conclusions}\label{Sec:Conclusions}
We have studied the interaction of $\phi^6$ kinks with a $\mathcal{PT}$-symmetric defect in which gain and loss are balanced. An analytical collective variable method was used to compare the results obtained with the numerical simulations of continuous model.

In contrast to sG and $\phi^4$ models studied in \cite{Danial.PRE.2014,Danial.CNSNS.2015}, when $\phi^6$ kink comes from the gain side its final velocity decreases after passing through the defect, while for an antikink increasing of final velocity was observed (see Fig.~\ref{fig:FinalVelocity}~(a)). This effect of the kinks velocity shift increases for higher initial kink velocities. Surprisingly, kink and antikink have similar behaviour when they approach the defect from the loss side, i.e., the final velocity of the kinks does not change and the scattering is almost elastic (see Fig.~\ref{fig:FinalVelocity}~(b)). This can be explain by the fact that the energy transmission between kink's internal structure and translational mode of the kink is possible when the kink comes from the gain side (see Fig.~\ref{fig:ElasticStrain}).

A kink-defect interaction in $\phi^6$ model is more interesting due the antisymmetric nature of the kink in this model. It was demonstrated that when the kink (and antikink) hits the defect from the gain side multiple pairs of the kink-antikink pairs are formed in the final states depending on the kink initial velocity and also amplitude of the defect (see Fig.~\ref{fig:3Dplot}). Due to conservation of topological charge, only even number of kinks, i.e., kink and antikink pairs were formed after the interaction with defect. The number of pairs created in the final states increases with increasing kink initial velocity and also amplitude of the defect (see Figs.~\ref{fig:NOK}~(a,b)).

A well-separated kink and antikink pair moving toward the defect from the both side is also considered in this paper. For $V<V_c$ the collision between kinks happens in the loss region and this lead to the formation of phonons in the final states. However, for $V>V_c$ kink and antikink may interact in the gain region and formed the second pair of the kink-antikink or/and a localized bound state, bion (see Fig.~\ref{fig:2kcollision}).

We conclude that in the presence of $\mathcal{PT}$-symmetric defects, phonons can acquire energy from the defect (in particular from the gain region) and transform into multiple pairs of soliton and anti-soliton which may have many physical applications for future studies.

Finally, we would like to mention that the interaction of kinks with a $\mathcal{PT}$-symmetric defect in higher order of filed theories give new opportunities in the utilization of the soliton dynamics and open new directions for future investigations. Kinks in the $\phi^6$ model are asymmetric and have short-range tails, while in the $\phi^8$ and other higher-order models, e.g., $\phi^{10}$ and $\phi^{12}$ the kinks can have long-range tails with power-law decay. In addition to this, kinks in the $\phi^8$ model can also have internal vibrational modes depending on the parameters of the model. These modes can affect the energy redistribution into the phonons in the presence of the $\mathcal{PT}$-symmetric defect and make the dynamic of the system more interesting. Such studies are currently under investigation and will be reported in future publications.

\section*{Acknowledgments}
For A.M.M., this work is supported by Islamic Azad University Quchan branch under the grant.

\end{document}